\documentclass[%
twocolumn,
amsmath,amssymb,amsfonts
]{revtex4-1}

\usepackage{graphicx}% Include figure files
\usepackage{dcolumn}% Align table columns on decimal point
\usepackage{bm}% bold math
\usepackage[all]{xy}

\usepackage{xcolor}
\definecolor{darkblue}{rgb}{0.1,0.1,.7}
\usepackage[colorlinks, linkcolor=darkblue, citecolor=darkblue, urlcolor=darkblue]{hyperref} 
\usepackage{dsfont}

\usepackage{subcaption}
\usepackage[utf8]{inputenc}
\usepackage[vcentermath]{youngtab}

\usepackage{setspace}

\usepackage{amsmath, amssymb, amsthm, float, graphicx}
\usepackage{booktabs}
\usepackage{enumitem}

\def\beq{\begin{eqnarray}}\def\eeq{\end{eqnarray}}
\def\be{\begin{equation}}\def\ee{\end{equation}}

\def\eps{\epsilon}

\def\g{\gamma}

\def\s{\sigma}
\def\m{\mu}

\def\a{\alpha}
\def\e{\epsilon}

\def\b{\beta}
\def\d{\delta}
\def\c{\chi}

\def\vf{\varphi}

\def\D{\Delta}

\def\l{\lambda}

\def\la{\langle}
\def\ra{\rangle}

\def\Ocal{{\mathcal{O}}}
\def\Scal{{\mathcal{S}}}
\def\Jcal{{\mathcal{J}}}
\def\Bcal{{\mathcal{B}}}

\def\Mcal{{\mathcal{M}}}
\def\Lcal{{\mathcal{L}}}

\def\tr{{\rm tr~}}

\def\thetab{{\bar{\theta}}}
\def\thetabar{{\bar{\theta}}}
\def\psib{\bar{\psi}}
\def\psibar{\bar{\psi}}

\newcommand{\tmop}[1]{\ensuremath{\operatorname{#1}}}

\begin{document}
\vspace*{-.6in}
		 \begin{flushright}
 CPHT-RR113.122021 
  \end{flushright}
	\preprint{APS/123-QED}
	%  \vspace*{-.6in}
	\title{The Fate of Parisi-Sourlas Supersymmetry in Random Field Models}% Force line breaks with \\
   %new title proposed

	\author{Apratim Kaviraj$^{a,b,c}$, Slava Rychkov$^{b,d}$ and Emilio Trevisani$^{b,e}$}

	\affiliation{ \ \\
		$^{a}$Institut de Physique Th\'{e}orique Philippe Meyer \& $^{b}$Laboratoire de Physique de l’Ecole normale sup\'erieure, ENS,\\
	{Universit\'e PSL, CNRS Sorbonne Universit\'e \text{$,$} Universit\'e de Paris \text{$,$} F-75005 Paris, France}\\
	$^{c}$DESY Hamburg, Theory Group\text{$,$}  Notkestra\ss e 85\text{$,$}  D-22607 Hamburg, Germany\\
	$^{d}$Institut des Hautes \'Etudes Scientifiques\text{$,$} Bures-sur-Yvette\text{$,$} France\\
		$^{e}$CPHT, CNRS, Ecole Polytechnique\text{$,$}  IP Paris\text{$,$}  F-91128 Palaiseau, France
	}%

	\begin{abstract}
		
%The Parisi-Sourlas conjecture says that the	critical point of a random field (RF) theory is described by  a supersymmeric (SUSY) CFT	and related to a $d-2$ dimensional CFT. Numerical studies suggest that this is true for an RF $\phi^3$ model but not for RF $\phi^4$ in $d<5$. We discuss from an axiomatic CFT viewpoint how the SUSY CFT$_d$ dimensionally reduces to a CFT$_{d-2}$. Then we show how a SUSY fixed point emerges in an RF model from a sector of operators (`leaders') that controls the  RG flow. From a  classification of all leaders we determine the stability of this fixed point. Our predictions are in good agreement with the Mone Carlo findings for both the above RF models.  

By the Parisi-Sourlas conjecture, the critical point of a theory with random field (RF) disorder is described by a supersymmeric (SUSY) conformal field theory (CFT), related to a $d-2$ dimensional CFT without SUSY. Numerical studies indicate that this is true for the RF $\phi^3$ model but not for RF $\phi^4$ model in $d<5$ dimensions. Here we argue that the SUSY fixed point is not reached because of new relevant SUSY-breaking interactions. We use perturbative renormalization group in a judiciously chosen field basis, allowing systematic exploration of the space of interactions. Our computations agree with the numerical results for both cubic and quartic potential.
	\end{abstract}
	
	%\keywords{Suggested keywords}%Use showkeys class option if keyword
	%display desired
	\maketitle

%	\tableofcontents
	
%	\section{\label{sec:level1}Introduction}
	%\ \\
	%{\large \bf Introduction.} \ 
	
\emph{Introduction} --- Emergent symmetries are a frequent theme in modern theoretical physics. Such a symmetry is present at long distances but is not visible in the microscopic description of the system. A beautiful example is furnished by the physics of disordered systems, namely by the Random Field Ising Model (RFIM) and its cousins. Parisi and Sourlas suggested long ago \cite{Parisi:1979ka,PhysRevLett.46.871} that the critical points of these models obey emergent supersymmetry. While supersymmetry plays a prominent role in high-energy physics, its appearance in the statistical physics context came as a major surprise. A dramatic consequence of supersymmetry is dimensional reduction \cite{Aharony:1976jx}: the critical exponents of a disordered system in $d$ dimensions should be the same as those of the pure (i.e.~non-disordered) system in $d-2$ dimensions. 

Unfortunately, after 40 years of work, there is still no complete understanding whether, when, and how Parisi-Sourlas supersymmetry actually emerges. Most work focused on the random field $\phi^4$ and $\phi^3$ field theories, describing respectively the phase transition in RFIM and the statistics of Branched Polymers (BP) in a solution \cite{PhysRevA.20.2130,Redner_1979,Gaunt_1980}. Numerical studies of microscopic models suggest that supersymmetry and dimensional reduction are present in any dimension for the $\phi^3$ case \cite{LatticeAnimals} but only in sufficiently high $d$ for the $\phi^4$ case \cite{Fytas3,Picco1,Picco2,Picco3}. Why does this happen? One possibility is that some SUSY-breaking perturbations are \emph{dangerously irrelevant}, i.e.~irrelevant for high $d$, while become relevant at lower $d$ and break supersymmetry \cite{Brezin1998,Feldman} \footnote{We stress that such a SUSY-breaking is explicit and not spontaneous.}. In this Letter we will report the first systematic exploration of this scenario. We will show that it gives a satisfactory unified description of phenomenology in agreement with all available numerical results \footnote{Other theoretical ideas and methods used for understanding the phase transition in the RF models and the loss of PS SUSY include: formation of bound state of replicas \cite{Brezin2001,PSbound}, expansion at high temperature \cite{HighT} and around the Bethe lattice \cite{Parisi2019}, and the conformal bootstrap \cite{Hikami:2017sbg,Hikami:2018mrf}. Comparison to functional renormalization group studies will be given at the end.}.

\emph{The model and prior work} ---
A random field (RF) model describes a statistical field theory with quenched disorder coupled to a local order parameter. We consider RF models of the type
\be\label{eq:RFdef}
\mathcal{S}[\phi,h]=\int d^d x\Big[\frac{1}{2}(\partial_\mu\phi)^2+V(\phi)+h(x)\phi(x)\Big]\,,
\ee
where $h(x)$ is drawn from a Gaussian distribution with zero mean and $\overline{h(x)h(0)}=H\delta(x)$\,.
%Critical points of these theories were hypothesized to be related to the critical point of the same theory without disorder in two less dimensions $\widehat{d}\equiv d-2$ \cite{Aharony:1976jx}. 
Parisi-Sourlas (PS) conjecture \cite{Parisi:1979ka} about the critical points of these theories can be naturally divided in two parts:
%Part 1 ({\it Emergence of SUSY}) says that the critical point of an RF theory is a SUSY CFT of a special kind (PS CFT). Part 2 ({\it Dimensional reduction}) identifies a large class of observables of the PS CFT (e.g.~its critical exponents) described by a CFT in two less dimensions. 

\noindent\begin{enumerate}[wide]%[topsep=0.1 cm, itemsep=0.1 cm, partopsep=0 cm, parsep=0 cm]
	\item {\it Emergence of SUSY}: The critical point of an RF theory is described by a special SUSY CFT (PS CFT).
	
	\item  {\it Dimensional reduction}: A large class of observables of the PS CFT (e.g. its critical exponents) are described by an ordinary CFT living in $\widehat{d}\equiv d-2$ dimensions.
\end{enumerate}

While perturbatively valid for $d$ infinitesimally close to the upper critical dimension $d_{uc}$ (see below), this remarkable conjecture is known to sometimes fail for the physically interesting cases of integer $d<d_{uc}$. %The purpose of this letter is to explain why this happens.

As mentioned, the two most studied RF models are with $\phi^4$ (RFIM) and $\phi^3$ (BP) potentials.
%This remarkable conjecture is known to fail in some cases (see below) and while being studied for decades, there is still no consensus on why this happens.	
The RF $\phi^4$ model has a critical point in $3 \le d < d_{uc}=6$.
PS conjecture would relate it to the usual Ising model in $\widehat{d}$ dimensions.
Numerical studies \cite{Fytas3,Picco1,Picco2,Picco3} show that while both SUSY and dimensional reduction hold in $d=5$, the conjecture fails in $d=4$. It also fails trivially for $d=3$, as the $\widehat{d}=1$ Ising model has no phase transition. 

Similarly, the critical point of the RF $\phi^3$ model with imaginary coupling should be described by the usual Lee-Yang fixed point in $\widehat{d}$ dimensions \cite{Fisher:1978pf}.
BP critical exponent simulations suggest that this instance of PS conjecture works perfectly for any 
$2\le d< d_{uc}= 8$ \cite{LatticeAnimals} \footnote{A special model of BP with microscopically realized SUSY \cite{zbMATH02068689} was proven to undergo dimensional reduction in any $d$. This result does not apply to generic (non SUSY) BP models or to the RF $\phi^3$ itself and does not shed light on why PS conjecture works for those models.}.

Let us come back to the central question of why PS conjecture sometimes works and sometimes fails. Many perturbative and non-perturbative arguments were given for Part 2 of the conjecture \cite{Parisi:1979ka,CARDY1983470,KLEIN1983473,Klein:1984ff,Zaboronsky:1996qn,paper1}. On the other hand Part 1 appears to be on less solid grounds. Here we will focus on the scenario \cite{Brezin1998,Feldman} that Part 1 may fail due to dangerously invariant SUSY-breaking interactions. %Here we will present the first systematic exploration of this scenario. %We will see that it provides a unified explanation for PS SUSY presence or absence in BP and Ising models across dimensions.

\emph{From replicas to Cardy fields} --- We start by using the usual replica method where we take $n$ copies of the action \eqref{eq:RFdef} and average out the disorder. This gives the replica action:
\be\label{eq:replicaaction}
\mathcal{S}_n=\int  \! d^dx \Big[\sum_{i=1}^n\big[(\partial_\mu\phi_i)^2+V(\phi_i)\big]-\frac{H}{2}\big(\sum_{i=1}^n\phi_i\big)^2\Big]
\ee 
from which  one can get quenched averaged correlations functions $\overline{{\langle A(\phi)\rangle}}$ in $n\to 0$ limit by simply computing ${{\langle A(\phi_1)\rangle}}$ having a single replica field. 

We next apply Cardy's linear field transform \cite{CARDY1985123}:
\be
\label{Ctrans}
\varphi=\frac{1}{2}(\phi_1+\rho)\,, \ \ \omega=\phi_1-\rho\,, \ \ \chi_i\stackrel{i\ne 1}{=}\phi_i-\rho\,,
\ee
with $\rho=\frac{1}{n-1}\sum_{i=2}^n \phi_i$ and the condition $\sum_{i=2}^{n}\chi_i=0$. Turning off interactions for now ($V=0$), the transformed Lagrangian takes the form
\be\label{eq:Cardylag}
\Lcal^{\text{free}}=\partial_\mu \vf \partial_\mu \omega-\frac H2 \omega^2 +\frac 12 \sum_{i=2}^n(\partial_\mu \chi_i)^2\,.
\ee
Here and below, because of the replica limit $n\to 0$, we are dropping all terms proportional to powers of $n$.% \footnote{The subtleties of the $n\to 0$ limit are discussed in \cite{paper2}.}.

From \eqref{eq:Cardylag} we read off the classical scaling dimensions of the Cardy fields: $[\vf]=\frac d2-2, [\chi_i]=\frac d2-1,[\omega]=\frac d2$\,. In contrast, the original replica fields $\phi_i$ do not even have a well-defined scaling dimension \footnote{This is clear e.g.~from their propagator mixing different powers of momentum, see \cite{Cardy-book}, Eq.~(8.39).}. Although not manifest in the Cardy field basis, the S$_n$ symmetry is still present and in particular not spontaneously broken \footnote{We note in this respect that replica symmetry breaking is proven not to happen in the RFIM \cite{Chatterjee}.}. It will play an important role below.

While RF criticality is often described in terms of special ``zero-temperature fixed points'' \cite{Bray1985, Fisher86}, Cardy transform puts it on the same footing as the more familiar non-disordered criticality. 
Using Cardy fields, we will be able perform the RG analysis for the RF models borrowing the standard Wilsonian methodology \cite{Wilson:1973jj,Kleinert:2001ax}. 

%		\be\label{eq:dim}
%	[\vf]=\frac d2-2, [\chi_i]=\frac d2-1,[\omega]=\frac d2\,.
%	\ee

% (see App.\ref{}).
%To arrive at \eqref{eq:Cardylag} we discarded operators that are proportional to powers of $n$.%  At $n\to 0$ we indeed get the fixed point defined by $\Lcal_0^{\text{free}}$. 

\emph{Leaders and followers} --- Let us now turn the interactions back on, and see how the theory renormalizes. Lagrangian \eqref{eq:replicaaction} contains the interaction $\sum_{i=1}^n V(\phi_i)$. This can be written as a sum of basic S$_n$ singlet interactions $\sigma_{k}\equiv \sum_{i=1}^n \phi_i^{k}$. In an exhaustive analysis, we will have to consider further interaction terms respecting the replica permutation symmetry S$_n$, since they will be generated by RG evolution \cite{Brezin1998}. Examples of such allowed interactions are products of $\sigma_{k}$'s as well as interactions containing derivatives. We will classify S$_n$ singlet interactions in the original fields of \eqref{eq:replicaaction}, and then transform them to the Cardy fields.% The simplest examples of such allowed S$_n$ singlet interactions are $\sigma_{k}\equiv \sum_{i=1}^n \phi_i^{k}$ and $\sigma_{k(\mu)(\mu)}\equiv \sum_{i=1}^n \phi_i^{k-2}\partial_{\mu}\phi_i\partial_\mu\phi_i$, and products of such terms \cite{Br_zin_1998}. 

%The free part of \eqref{eq:replicaaction} is composed of $\sigma_{2(\mu)(\mu)}$ and $\sigma_1^2$ which in Cardy variables reproduce $\Lcal^{\text{free}}$. 
The simplest interaction is the mass term $\sigma_2$ which in Cardy fields reads $2\vf\omega + \chi_i^2$ and has classical dimension $d-2$. Continuing at the cubic level, the operator $\sigma_3$ under Cardy transform becomes
\be
\sigma_3=(3\vf^2\omega + 3\chi_i^2\vf)+(\chi_i^3)-\big(\frac 32 \chi_i^2\omega\big)+\big(\frac 14 \omega^3\big)\,,
\ee
where different terms have unequal classical dimensions: $\frac{3d}{2}-4$ for the first term, while the successive ones sit 1,2 and 3 units higher.
%	where classical dimension of each term in bracket is one higher than the preceding one \ET{this is actually false...}, with the lowest one being $\frac{3d}{2}-4$.
This new effect is generic: any singlet operator $\mathcal{O}$ in Cardy fields can be written as 
\be
\mathcal{O}=\Ocal_L+\Ocal_{F_1}+\Ocal_{F_2}+\cdots\,,
\ee
where $[\mathcal{O}_{F_i}]=[\mathcal{O}_L]+i$, $i=1,2,\ldots$. We call the lowest dimension part $\mathcal{O}_L$ the `leader', and $\mathcal{O}_{F_i}$ `followers'. 

In the first part of a Wilsonian RG step, integrating out a momentum shell and lowering the momentum cutoff $\Lambda\to \Lambda/b$ ($b>1$), a singlet operator $\mathcal{O}$, if present in the effective action, renormalizes as a whole, i.e.~only through the change of the overall coefficient: $g \mathcal{O} \to \tilde g \mathcal{O}$ \footnote{Here for simplicity we ignore interaction mixing effects, taken into account in the computations described below.}. This is guaranteed by S$_n$ symmetry. %which, although non-manifest, is present even when working in Cardy fields. 
On the other hand, in the second part of an RG step, bringing the cutoff back up to its original value, which rescales the fields $\vf,\chi_i,\omega$ according to their classical dimensions, the followers rescale by different coefficients from the leader, suppressing their relative effect in the IR (i.e.~at large $b$):
\be
\Ocal_L+\sum\nolimits_i \Ocal_{F_i} \to b^{-[\Ocal_L]}\big(\Ocal_L+\sum\nolimits_i{b^{-i}}\Ocal_{F_i}\big).
\ee
Hence, the RG flow in the IR is controlled by the leaders. This drastically reduces the number of interactions to consider: only operators in Cardy fields which can be written as a leader of an S$_n$ singlet interaction are of interest. The RG relevance or irrelevance of the leader determines the fate of the whole interaction \cite{paper2}.

%	\bf Emergence of SUSY at $\e\ll1$.} \ 
Keeping the free massless Lagrangian \eqref{eq:Cardylag}, the mass term, and the leader parts $(\s_2)_L$ or $(\s_3)_L$ of the $\phi^3$ or $\phi^4$  interactions, we get the two Lagrangians relevant for the description of the RF $\phi^3$ and $\phi^4$ models:
\begin{gather}
\Lcal^{\phi^3}_L=\Lcal^{\text{free}}+m^2(2\vf\omega + \chi_i^2)+ \frac{g}{2} (\vf^2\omega + \chi_i^2\vf)\,, \label{Leff}
\\
\Lcal^{\phi^4}_L=\Lcal^{\text{free}}+m^2(2\vf\omega + \chi_i^2)+ \frac{g}{12} (2\vf^3\omega +3 \chi_i^2\vf^2)\, .\nonumber
\end{gather}
The mass term $m^2$ is strongly relevant and should be tuned to reach the IR fixed point. The upper critical dimension in this approach is fixed simply from the marginality of the leading non-quadratic interaction, which gives the well-known values cited above: $d_{uc}=8$ for the $\phi^3$ and 6 for the $\phi^4$ models. 

Eqs.~\eqref{Leff} give the correct effective theory for the two models close to their upper critical dimension, i.e.~for $d=d_{uc}-\e$, $\e\ll 1$. Indeed, one can check that in this case, no other S$_n$ singlet interactions exist whose leaders would be relevant (and, for the $\phi^4$ case, respecting the extra $\mathbb{Z}_2$ symmetry).  
However, we should keep an open mind about what may happen for $\e=O(1)$, as some irrelevant interactions may become relevant. This will be investigated below.

\emph{Emergence of SUSY} --- It is easy to see that both Lagrangians \eqref{Leff} have emergent SUSY \cite{CARDY1985123}.
Note that the $n-2$ fields $\chi_i$ appear quadratically in the Lagrangians. The associated partition function is given by a Gaussian integral which at $n\to 0$ is equal to that of $2$ anticommuting scalars $\psi,\psib$. So we are allowed to replace $\chi_i \chi _i \to 2 \psi \psib$. Then both the above theories can be compactly written as % (for the cubic theory $V(\phi)=m^2\phi^2+\frac{g}{6}\phi^3$, while for the quartic $V(\phi)=m^2\phi^2+\frac{\l}{4!}\phi^4$) :
\be \label{eq:susy}
\mathcal{S}_{\text{susy}}=\int d^dx d\theta d\thetab \Big[-\frac 12 \Phi  \partial^a\partial_a \Phi+V(\Phi)\Big]\, ,
\ee
where  $V(\phi)=m^2\phi^2+\frac{g}{6}\phi^3$ for the cubic theory and $V(\phi)=m^2\phi^2+\frac{\l}{4!}\phi^4$ for the quartic.
%Here $D^2=\partial^2-H\partial_\theta\partial_{\thetab}$ and $\Phi(x,\theta,\thetab)=\vf +\theta \psib+\thetab \psi+\theta\thetab \omega$\,. 
Here $\Phi(x,\theta,\thetab)=\vf +\theta \psib+\thetab \psi+\theta\thetab \omega$ is a superfield depending on coordinates $x,\theta,\thetab$ parametrizing the superspace $\mathbb{R}^{d|2}$ with OSp$(d|2)$ supergroup symmetry (PS supersymmetry), and $ \partial^a\partial_a$ is the super-Laplacian (index $a$ takes values $1,\ldots,d,\theta,\thetab$). In the IR, we get a further enhancement to a PS superconformal symmetry OSp$(d+1,1|2)$ \cite{KUPIAINEN1983380}. The fixed point of this theory is therefore a PS CFT.

We now briefly describe basic properties of PS CFTs and how they undergo dimensional reduction
% \footnote{\label{Appendix}See appendices for more details on PS SUSY, dimensional reduction, properties of leader operators, and RG computations presented in the main text.}. 
\cite{appendix}
Local operators in such theories are classified according to their superconformal dimension $\D$ %(measured by the dilatation generator $D$ of the superconformal algebra) 
and their OSp$(d|2)$ spin $\ell$. They are grouped in superconformal multiplets containing a superprimary operator $\mathcal{O}^{\bf a}_{\D \ell}$ (where ${\bf{a}}$ stands for $a_1 a_2 \dots$), annihilated by the special superconformal generator $K^a$, and its superdescendants such as $\partial_a \mathcal{O}^{\bf a}_{\D \ell}$ and higher superderivatives.
%Denoting $\{a_\ell\}=a_1\cdots a_\ell$, a superprimary ${\mathcal{O}^{\{a_\ell\}}}(y)$ decomposes into conformal primaries $\mathcal{O}^{\{a_\ell\}}(x)$ as follows:
$\mathcal{O}^{\bf a}_{\D\ell}$ can be expanded in components which have different conformal dimensions:
\be\label{supfieldexp}
{\mathcal{O}}^{\bf{a}}(y)=
\underbrace{\mathcal{O}_{0}^{\bf{a}}(x)}_{\D}+
\theta \underbrace{ \mathcal{O}_{\theta}^{\bf{a}}(x)}_{\D+1}+
\bar{\theta} \underbrace{ \mathcal{O}_{\bar{\theta}}^{\bf{a}}(x)}_{\D+1}+
\theta\bar\theta \underbrace{ \mathcal{O}_{\theta\bar\theta}^{\bf{a}}(x)}_{\D+2}\, .
\ee 

Dimensional reduction restricts correlators of a PS CFT to a $(d-2)$-dimensional bosonic subspace $\mathcal{M}_{\widehat d}\equiv \{y=(\widehat x^\a,0,0,0,0), \widehat x \in \mathbb{R}^{\widehat d}\}$. In addition, one only considers PS CFT operators invariant under the subgroup OSp$(2|2)$ (super)rotating the directions orthogonal to $\mathcal{M}_{\widehat d}$.
In general, restricting to a subspace gives a nonlocal theory. The nontrivial fact is that by restricting the OSp$(2|2)$-singlet sector of the SUSY theory, we get a local ${\widehat d}$-dimensional CFT living on $\mathcal{M}_{\widehat d}$ \cite{paper1}. The local conserved CFT$_{\widehat d}$ stress tensor appears in this setup as the $\mathcal{T}_0$ component of the PS CFT superstress tensor $\mathcal{T}$. \label{dimredpage}

The dimensionally reduced CFT$_{\widehat d}$ has the global symmetry of the original PS CFT: trivial in the $\phi^3$ case and $\mathbb{Z}_2$ for $\phi^4$. We will naturally assume that this CFT$_{\widehat d}$ is nothing but the ${\widehat d}$-dimensional critical point of the same theory without disorder \footnote{This is also supported by perturbative \cite{Parisi:1979ka, paper2,paper3} and rigorous Lagrangian arguments \cite{KLEIN1983473,Klein:1984ff}.}: the Wilson-Fisher fixed point for $\phi^4$ \cite{Wilson:1971dc} and the Lee-Yang fixed point for $\phi^3$ \cite{Fisher:1978pf}. Dimensions of many operators in these familiar theories being well-known both perturbatively and, sometimes, non-perturbatively, we can then use dimensional reduction to infer dimensions of operators in the PS CFT. 

The central question is whether any S$_n$ singlet perturbation, while irrelevant for $\e\ll 1$, may become relevant for $\e=O(1)$ and destabilize the SUSY IR fixed point.
As discussed above, this may be answered by perturbing the Lagrangians $\Lcal_L$ in \eqref{Leff} by the leader terms of S$_n$ singlet interactions, and computing their scaling dimensions (restricting to $\mathbb{Z}_2$ singlets for $\phi^4$ case). A priori there are many leaders to consider, which moreover may mix under RG.
Below we will divide them into three classes: susy-writable (SW), susy-null (SN), and non-susy-writable (NSW), with a triangular mixing matrix. Namely SN operators can generate only SN under RG flow, SW can generate SW and SN, while NSW can generate all three classes. %We will now describe the three classes and the anomalous dimensions of low-lying operators in each class (Table \ref{tab:table3}).

\emph{Susy-writable (SW) leaders} ---
\label{SW}
These are invariant under $O(n-2)$ acting on the indices of the $\chi_i$ fields. These operators can be transformed to the SUSY field bases by the substitution $\c_i \to \psi$ (hence the name). With abuse of language we will also refer as SW to the resulting Sp$(2)$-invariant operators. In addition, we require that the operator does not vanish after the substitution (if so it will be classified below as susy-null). Most low-lying leaders turn out to be SW. E.g.~the leader of any S$_n$ singlet $\sum_{i = 1}^n A (\phi_i)$ has the form $ A'(\varphi) \omega 
+  \frac{1}{2}A''(\varphi)  \chi_i^2$ which is  SW. This can be written as the highest component $A_{\theta \thetab}(\Phi)$ of a scalar composite superfield $A(\Phi)$.
More generally, SW leaders are always in the highest component of a superfield \footnote{This was first conjectured in \cite{paper2} and now we have a rigorous proof \cite{appendix}.}. They do not have to be scalars of OSp($d|2$), but only singlets of the subgroup $\text{SO}(d) \times \text{Sp}(2)$. These are obtained from a highest component $ \mathcal{O}^{\bf a}_{\theta \thetab}$ by contracting all its ${\bf a}$ indices with the $\text{Sp}(2)$-metric i.e.~by setting the indices to $\theta$ and $\thetab$ \cite{appendix}. 

The OSp$(d|2)$ tensor representations of ${\mathcal{O}}^{\bf{a}}$ are associated to Young tableaux (YT) $(\ell_1,\ell_2,\cdots)$ with $\ell_i$ boxes in $i$-th row. Indices along the rows (columns) are graded (anti)symmetrized and all supertraces removed. Graded symmetry and antisymmetry respectively mean $\Ocal^{ab}=(-1)^{[a][b]}\Ocal^{ba}$ and $\Ocal^{ab}=-(-1)^{[a][b]}\Ocal^{ba}$ where $[a]=0(1)$ if $a$ is bosonic (fermionic). These general facts combined with the above procedure of setting the indices to $\theta$ and $\thetab$ shows that SW leaders can only be obtained from operators in representations labelled by YT of the form $(2,2,\dots,2)$. SW leaders are thus in correspondence with the following superfields $\Scal_{\theta \thetab},  \Jcal^{\theta \thetab}_{\theta \thetab},  \Bcal^{\theta \thetab, \theta \thetab}_{\theta \thetab}, \dots$
%	\begin{equation}
%	\Scal_{\theta \thetab} \, ,
%	\qquad 
%	\Jcal^{\theta \thetab}_{\theta \thetab} \, ,
%	\qquad 
%	\Bcal^{\theta \thetab, \theta \thetab}_{\theta \thetab} \, , 
%	\quad 
%	\dots  
%	\end{equation} 
where $\Scal$ is a scalar,  $\Jcal^{ab}$ a spin-two, and $\Bcal^{ab,cd}$ a ``box'' operator in the YT $(2,2)$ representation where $(a,b)$ and $(c,d)$ are the graded-symmetric pairs. Representations with higher number of rows can also appear in generic $d$ but we do not consider them since they have large classical dimensions. 

The above formal considerations have a neat practical consequence: dimensions of SW leaders $\mathcal{O}_{\theta\thetab}$ can be obtained by studying the respective operators $\widehat \Ocal$ in the dimensionally reduced model using $\D_{\mathcal{O}_{\theta\thetab}}=\D_{\mathcal{O}}+2 = \D_{\widehat \Ocal}+2$ from \eqref{supfieldexp}. From here we see immediately that SW leaders originating from scalar and spin two PS CFT operators cannot %play an important role in destabilizing
destabilize the SUSY fixed point. Indeed in both dimensionally reduced models all scalars (besides the mass term which we tune to reach the fixed point) are irrelevant. Similarly all the spin two operators should not cross the stress tensor and thus are expected to remain irrelevant in any $d$ \footnote{\label{T_redundant}The SW leader perturbation originating from the superstresstensor, $\mathcal{T}^{\theta \thetab}_{\theta \thetab}$, deserves a special comment. Naively it has dimension $d$ and is marginal. However, it is more properly classified as redundant. Its only effect is to rescale factor $H$ in \eqref{eq:Cardylag}, which is also a parameter entering the superspace metric. This rescaling can be undone by field redefinition and has no physical consequences.}.

Separate analysis is needed for operators in the box representation. In the dimensionally reduced models, an infinite family of such operators can be written in terms of $\widehat{d}$-dimensional scalar field $\widehat{\phi}$ as
\begin{equation}
\widehat{B}^{(k)}_{\a \b , \g \d} \equiv \widehat{\phi}^{k-3} \left( \widehat{\phi}_{, \a \b} \widehat{\phi}_{,
	\g \d} \widehat{\phi} - \tfrac{2 \widehat{d}}{\widehat{d} - 2} \widehat{\phi}_{, \a}
\widehat{\phi}_{, \b} \widehat{\phi}_{, \g \d}  \right)_Y \,, 
\label{Box-n}
\end{equation}
with $k\ge 3$. Greek letters denote $\mathbb{R}^{\widehat{d}}$ indices, and $Y$  indicates the box YT symmetrization, the two symmetric rows being $\a \b$ and $\g \d$. These are the lowest dimensional operators made of $k$ fields in such representation. 

We computed their perturbative one-loop dimensions for the $\phi^3$ case \cite{paper3}, following the standard $\e$-expansion methodology \cite{Wilson:1973jj,Kleinert:2001ax}, while the $\phi^4$ case was considered previously in \cite{Kehrein:1994ff}. The results (classical dimension plus one-loop correction) are:
 \begin{gather}
 	\D_{\mathcal{B}_{\theta\thetab}^{(k)}}=\begin{cases}\big(2k+6-\frac{k}{2}\e\big)_{\text{cl}}+ \frac{ 2 k^2-5 k-2 }{6} \epsilon &(\phi^3)\,,\\
 	%\D_{\mathcal{B}_{\theta\thetab}^{(k)}}=
 \big(k+2-\frac{k }{2} \epsilon \big)_{\text{cl}}+
 \frac{(k-3) (3 k+2) }{18}\e &(\phi^4) \,.
 	\end{cases}
 \label{Bk}
 \end{gather}
Importantly, all anomalous dimensions are positive (excluding the $k=3$ $\phi^4$ case which, as all odd $k$ for $\phi^4$, is unimportant since it does not respect $\mathbb{Z}_2$ symmetry).

 %\cite{Kehrein:1992fn,paper2,paper3}. %Liendo:2017wsn

\emph{Susy-null (SN) leaders} ---%\label{sec:susynull}
\label{susynull}
%%%%%%%%%%%%%%%%%%
These are  singlets under $O(n-2)$ (like the SW operators) and satisfy the property of vanishing under the $\chi\to\psi$ map by the Grassmann nature of $\psi$. A typical example is $(\chi_i^2)^2 \to (\psi\psib)^2=0$. 
These operators have restrictive mixing properties and can only generate operators of the same class under RG.
We identified an infinite class of S$_n$ singlets \cite{appendix}
\be
\label{def:NkSingl}
\mathcal{N}_{k}=\frac{2}{k-3} \left(\frac{\sigma _2 \sigma _{k-2}}{k-2}-\frac{2 \sigma _1 \sigma _{k-1}}{k-1}\right)  \, ,
\ee
for $k=4,5,6, \dots$, which have SN leaders $(\mathcal{N}_{k})_L=\varphi^{k-4}(\chi_i^2)^2$.
The $k=4$ operator is the lowest dimensional SN leader overall, while $(\mathcal{N}_{k})_L$ is the lowest dimensional SN leader made of $k$ fields. 

Unlike for SW leaders, we cannot use SUSY theory and dimensional reduction to infer the scaling dimensions of SN operators (since they vanish identically in SUSY fields). We compute them directly from action \eqref{Leff}. Our Cardy field approach makes these computations methodologically straightforward, being analogous to the standard $\e$-expansion \cite{Wilson:1973jj,Kleinert:2001ax}. We thus computed the leading anomalous dimension of operators \eqref{def:NkSingl}. The resulting scaling dimensions (classical plus one-loop) are given by:
\be
\D_{{(\mathcal{N}_{k})}_L}=\begin{cases}
	\big(2 (k+2)-\frac{\epsilon}{2} k \big)_{\textrm{cl}}
+\frac{ 6 k^2-7k-48}{18} \epsilon&(\phi^3),
\\
\big(k+4-\frac{\epsilon}{2}k  \big)_{\textrm{cl}} 
+\frac{(k-4) (k+3)}{6}\eps&(\phi^4).
\end{cases}
\label{Nk3}
\ee
The one-loop correction is positive except for the $k=4$, $\phi^4$ case when it vanishes. Then, the first nonzero correction appears at two loops, and it is negative \cite{paper2}:
\be
\D_{{(\mathcal{N}_{4})}_L} = (8-2\epsilon  )_{\textrm{cl}}
	-\frac{8}{27}\e^2\quad (\phi^4)\,.
	\label{N04}
\ee

\emph{Non-susy-writable (NSW) leaders} ---
%%%%%%%%%%%%%%%%%%
These operators are singlets under the S$_{n-1}$ that permutes the fields $\chi_i$, but not under $O(n-2)$, and therefore they cannot be mapped to $\psi, \psib$ fields. A typical example would be any leader involving $\sum_{i=2}^n\chi_i^3$. In the RG flow, leader perturbations belonging to this class can generate perturbations from the other two classes, while the opposite mixing is forbidden by SUSY.

We investigated two infinite families of S$_n$ singlets having NSW leaders \cite{appendix}. 
The first family, first discussed by Feldman \cite{Feldman} and in \cite{paper2}, is given by
\begin{equation}
	\mathcal{F}_k  =  \sum_{i, j = 1}^n (\phi_i - \phi_j)^k = \sum_{l = 1}^{k
		- 1} (- 1)^l \binom{k}{l} \sigma_l \sigma_{k - l} \,,  \label{Fk}
\end{equation}
with $k=6,8,10, \dots$ \footnote{For $k=2$ and $k$ odd we have $\mathcal{F}_{k} = 0$, while $\mathcal{F}_{4}\propto \mathcal{N}_4$ has a SN leader.}. They give rise to NSW leaders made only of $\chi$ fields, of the form
\begin{equation}
	\left( \mathcal{F}_k\right)_L  =  \sum_{l = 2}^{k - 2} (- 1)^l \binom{k}{l} (   \chi^l_i)  \left(  \chi^{k - l}_j \right) 
	\, ,
\end{equation}
The first leader of this family, $(\mathcal{F}_6)_L$, is the lowest dimensional NSW leader overall.

The second family consists of S$_n$ singlets given by
\begin{equation}
\mathcal{G}_{k}\equiv \frac{\sigma _3 \sigma _{k-3}}{3 (k-5)}+\frac{\sigma _1 \sigma _{k-1}}{k-1}-\frac{(k-4) \sigma _2 \sigma _{k-2}}{(k-5) (k-2)} \, ,
\end{equation}
for $k=6,7, 8, \dots$. These have NSW leaders 
\be
(\mathcal{G}_{k})_L ={\textstyle \frac{(k-4) (k-3)}{36}}
\vf^{k-6}\bigl [2 (\chi_i^3)^2 - 3 (\chi_i^2)(\chi_j^4)\bigr],
\ee
%with $C_k=\frac{(k-4) (k-3)}{36}$. 
The two families start from the same operator ($\mathcal{G}_{6}\propto \mathcal{F}_{6}$), but the higher operators are different. In fact $(\mathcal{G}_{k})_L$ is the lowest NSW leader made of $k$ fields, and in particular sits lower than $(\mathcal{F}_{k})_L$ for $k>6$.

%We now report our perturbative results for the scaling dimensions of these NSW leaders. 
Like for the SN class, we computed NSW scaling dimensions by the $\eps$-expansion methodology adapted to action \eqref{Leff}. 
Starting with the $\mathcal{F}_k$ family, the scaling dimension (classical plus the leading correction) is given by
\be
\D_{{(\mathcal{F}_{k})}_L}=\begin{cases}
	\big(3 k-\frac{k }{2}\e \big)_{\textrm{cl}}+
	\frac{ 2 k^2-3k}{18}\e &(\phi^3), \\
	\big(2 k - \frac{k}{2}\e \big)_{\textrm{cl}} 
	- \frac{k(3 k - 4) }{108}\e^2 &(\phi^4). 
\end{cases}
\label{Fk4}
\ee
Notably, the leading anomalous dimension is one-loop and positive in the $\phi^3$ case \cite{paper3} while it is two-loop and negative for $\phi^4$ \cite{Feldman,paper2}. 

Considering next the $\mathcal{G}_{k}$ family, we obtained
\be
\D_{{(\mathcal{G}_{k})}_L}=\begin{cases}
	\big( 2 (k+3)-\frac{ k }{2} \epsilon \big)_{\textrm{cl}}
+\frac{ 6 k^2-7k-120}{18} \epsilon&(\phi^3)
\\
\big(k+6 -\frac{k}{2}  \epsilon   \big)_{\textrm{cl}} 
	+\frac{  (k-6)(k+5)}{6} \epsilon\,.
	&(\phi^4)
	\end{cases}
	\label{Gk4}
\ee
The one-loop correction is therefore always positive, except in the $k=6$, $\phi^4$ case when it vanishes. In the latter case, using $\mathcal{G}_6\propto \mathcal{F}_6$ and Eq.~\eqref{Fk4}, we see that the leading, negative, correction appears at two loops.

\emph{Does SUSY emerge at $\e=O(1)$?} --- The analysis leading to SUSY was based on the effective Lagrangians \eqref{Leff}. It would be invalidated if a new relevant leader interaction is found in the IR. Allowed by symmetry, such a growing perturbation will be generated by the RG, destabilizing the flow and leading it away from the SUSY fixed point. Let us see if this scenario is realized.

Above we discussed several infinite families of leader interactions from three different classes (SW, SN, NSW). We will now focus on the lowest dimensional operators for each class. We expect them to be most important to decide the stability of the SUSY fixed point. First of all, $\e$-expansion computations of lowest-dimensional operators should be more reliable than for higher-dimensional ones \footnote{Loop corrections grow rapidly with the number of fields inside the operator, making naive extrapolation to $\e=O(1)$ questionable. See \cite{Badel:2019oxl} for related recent work.}. Second, we expect crossing of operator dimensions (within the same mixing class) to be avoided nonperturbatively. 

With this in mind we find that the SUSY IR fixed point of the RF $\phi^3$ theory should always be stable, since the lowest leader perturbations $\mathcal{B}_{\theta\thetab}^{(3)}$, $\mathcal{N}_{4}$, $\mathcal{F}_{6}$ never become relevant. To see this we take their one-loop dimensions given in Eqs.~\eqref{Bk},\eqref{Nk3},\eqref{Fk4} and use these expressions in the full range of interest $2\le d<8$ \footnote{Actually none of the infinitely many discussed leaders become relevant for the $\phi^3$ case.}.

However, the same argument for the $\phi^4$ case reaches a different conclusion \cite{paper2}.
While $\mathcal{B}_{\theta\thetab}^{(4)}$ remains irrelevant \footnote{If relevant, this operator would break SUSY since it breaks superrotations.}, both $(\mathcal{N}_{4})_L$ and $(\mathcal{F}_{6})_L$ become relevant at some critical dimension $d_c$ between four and five, namely $\Delta_{(\mathcal{N}_{4})_L} = d $  at $d= d_c \approx 4.6$ while $ \Delta_{(\mathcal{F}_{6})_L} = d $ when $ d_c \approx 4.2$. The precise value of $d_c$, and which of the two operators crosses marginality first, should be taken with a grain of salt coming from a two-loop computation. We may estimate the uncertainty replacing the expressions in Eqs.~\eqref{N04},\eqref{Fk4} by their $\textrm{Pad\'e}_{[1,1]}$ rational approximants. We then find that $(\mathcal{N}_{4})_L$ crosses marginality at $d_c \approx 4.7$, while $(\mathcal{F}_6)_L$ at $d_c \approx 4.5$. 

NSW interaction $(\mathcal{F}_{6})_L$ clearly breaks SUSY. Operator $(\mathcal{N}_{4})_L$ is also potentially SUSY-breaking, by affecting NSW coupling evolution (while being SN it does not directly affect SW sector).
We thus conclude that SUSY will be present in the RF $\phi^4$ model for $d_c<d<6$, while it will be lost for $d<d_c$ \footnote{Feldman \cite{Feldman} argued that SUSY will be lost arbitrarily close to $d=6$, because the negative anomalous dimension of interactions $\mathcal{F}_k$, $k\ge 8$, grows with $k$ making them to cross marginality closer and closer to $d=6$ as $k\to\infty$. We disagree with this argument as it does not take into account nonperturbative mixing \cite{paper2}. Our new results for the $\mathcal{G}_k$ family strengthen this objection. The second $\mathbb{Z}_2$ respecting operator of this family, $(\mathcal{G}_8)_L$, stays irrelevant due to its positive one-loop dimension (see \eqref{Gk4}). Nonperturbatively (while not in perturbation theory) $(\mathcal{G}_8)_L$ is expected to mix with Feldman operators, and level crossing will be avoided. Hence, we expect that $(\mathcal{G}_8)_L$ will provide a barrier which $(\mathcal{F}_k)_L$, $k\ge 8$, cannot cross, remaining irrelevant and unimportant for deciding the fate of SUSY.}.

Remarkably, our findings exactly match the expectations from numerical studies mentioned at the beginning, for both universality classes. %mentioned in the introduction, namely that the RF $\phi^3$ always has a SUSY IR fixed point, while the RF $\phi^4$ only has a SUSY fixed point for $d=5$, whereas for $d=3,4$ SUSY is lost. 
It is encouraging that already the leading order $\eps$-expansion results lead to this agreement. In the future, it would be interesting to determine our $d_c$ more accurately. This can be done systematically, increasing the perturbative order and using Borel resummation techniques, as is standard for the usual Wilson-Fisher fixed point \cite{Guida:1998bx,Kleinert:2001ax,Kompaniets:2017yct,Kompaniets:2019zes}. 

Finally, we wish to compare our results to functional renormalization group studies of the RF $\phi^4$ model, which also predict the loss of SUSY for $d<d_c^{FRG}\approx 5.1$ \cite{TarjusIV}. While their $d_c$ is similar, their mechanism is quite different from ours, being attributed to fixed point annihilation \cite{2020PhRvE.102f2154B}, so that below $d_c$ the SUSY fixed point does not exist. On the contrary, our SUSY fixed point exists for any $d$, being simply RG unstable for $d<d_c$. If so, one should be able to detect SUSY in lattice simulations for $d=4$, by performing additional tuning \footnote{But not for $d=3$, because in this dimension the SUSY fixed point ceases to exist, see \cite{paper2}, Section 3.1.}. This would be a decisive confirmation for our scenario.

A.K. is supported by DFG (EXC 2121: Quantum Universe, project 390833306), and E.T. by ERC (Horizon 2020 grant 852386). Simons Foundation grants 488655, 733758, and an MHI-ENS Chair also supported this work. We thank Kay Wiese for comments.

% The \nocite command causes all entries in a bibliography to be printed out
% whether or not they are actually referenced in the text. This is appropriate
% for the sample file to show the different styles of references, but authors
% most likely will not want to use it.
%\nocite{*}
% Produces the bibliography via BibTeX.
%\bibliographystyle{utphys}
\bibliography{mybib}

%merlin.mbs apsrev4-1.bst 2010-07-25 4.21a (PWD, AO, DPC) hacked
%Control: key (0)
%Control: author (8) initials jnrlst
%Control: editor formatted (1) identically to author
%Control: production of article title (-1) disabled
%Control: page (0) single
%Control: year (1) truncated
%Control: production of eprint (0) enabled
\begin{thebibliography}{62}%
\makeatletter
\providecommand \@ifxundefined [1]{%
 \@ifx{#1\undefined}
}%
\providecommand \@ifnum [1]{%
 \ifnum #1\expandafter \@firstoftwo
 \else \expandafter \@secondoftwo
 \fi
}%
\providecommand \@ifx [1]{%
 \ifx #1\expandafter \@firstoftwo
 \else \expandafter \@secondoftwo
 \fi
}%
\providecommand \natexlab [1]{#1}%
\providecommand \enquote  [1]{``#1''}%
\providecommand \bibnamefont  [1]{#1}%
\providecommand \bibfnamefont [1]{#1}%
\providecommand \citenamefont [1]{#1}%
\providecommand \href@noop [0]{\@secondoftwo}%
\providecommand \href [0]{\begingroup \@sanitize@url \@href}%
\providecommand \@href[1]{\@@startlink{#1}\@@href}%
\providecommand \@@href[1]{\endgroup#1\@@endlink}%
\providecommand \@sanitize@url [0]{\catcode `\\12\catcode `\$12\catcode
  `\&12\catcode `\#12\catcode `\^12\catcode `\_12\catcode `\%12\relax}%
\providecommand \@@startlink[1]{}%
\providecommand \@@endlink[0]{}%
\providecommand \url  [0]{\begingroup\@sanitize@url \@url }%
\providecommand \@url [1]{\endgroup\@href {#1}{\urlprefix }}%
\providecommand \urlprefix  [0]{URL }%
\providecommand \Eprint [0]{\href }%
\providecommand \doibase [0]{http://dx.doi.org/}%
\providecommand \selectlanguage [0]{\@gobble}%
\providecommand \bibinfo  [0]{\@secondoftwo}%
\providecommand \bibfield  [0]{\@secondoftwo}%
\providecommand \translation [1]{[#1]}%
\providecommand \BibitemOpen [0]{}%
\providecommand \bibitemStop [0]{}%
\providecommand \bibitemNoStop [0]{.\EOS\space}%
\providecommand \EOS [0]{\spacefactor3000\relax}%
\providecommand \BibitemShut  [1]{\csname bibitem#1\endcsname}%
\let\auto@bib@innerbib\@empty
%</preamble>
\bibitem [{\citenamefont {Parisi}\ and\ \citenamefont
  {Sourlas}(1979)}]{Parisi:1979ka}%
  \BibitemOpen
  \bibfield  {author} {\bibinfo {author} {\bibfnamefont {G.}~\bibnamefont
  {Parisi}}\ and\ \bibinfo {author} {\bibfnamefont {N.}~\bibnamefont
  {Sourlas}},\ }\href {\doibase 10.1103/PhysRevLett.43.744} {\bibfield
  {journal} {\bibinfo  {journal} {Phys. Rev. Lett.}\ }\textbf {\bibinfo
  {volume} {43}},\ \bibinfo {pages} {744} (\bibinfo {year} {1979})}\BibitemShut
  {NoStop}%
%%CITATION = PRLTA,43,744;%%
\bibitem [{\citenamefont {Parisi}\ and\ \citenamefont
  {Sourlas}(1981)}]{PhysRevLett.46.871}%
  \BibitemOpen
  \bibfield  {author} {\bibinfo {author} {\bibfnamefont {G.}~\bibnamefont
  {Parisi}}\ and\ \bibinfo {author} {\bibfnamefont {N.}~\bibnamefont
  {Sourlas}},\ }\href {\doibase 10.1103/PhysRevLett.46.871} {\bibfield
  {journal} {\bibinfo  {journal} {Phys. Rev. Lett.}\ }\textbf {\bibinfo
  {volume} {46}},\ \bibinfo {pages} {871} (\bibinfo {year} {1981})}\BibitemShut
  {NoStop}%
\bibitem [{\citenamefont {Aharony}\ \emph {et~al.}(1976)\citenamefont
  {Aharony}, \citenamefont {Imry},\ and\ \citenamefont {Ma}}]{Aharony:1976jx}%
  \BibitemOpen
  \bibfield  {author} {\bibinfo {author} {\bibfnamefont {A.}~\bibnamefont
  {Aharony}}, \bibinfo {author} {\bibfnamefont {Y.}~\bibnamefont {Imry}}, \
  and\ \bibinfo {author} {\bibfnamefont {S.~K.}\ \bibnamefont {Ma}},\ }\href
  {\doibase 10.1103/PhysRevLett.37.1364} {\bibfield  {journal} {\bibinfo
  {journal} {Phys. Rev. Lett.}\ }\textbf {\bibinfo {volume} {37}},\ \bibinfo
  {pages} {1364} (\bibinfo {year} {1976})}\BibitemShut {NoStop}%
%%CITATION = PRLTA,37,1364;%%
\bibitem [{\citenamefont {Lubensky}\ and\ \citenamefont
  {Isaacson}(1979)}]{PhysRevA.20.2130}%
  \BibitemOpen
  \bibfield  {author} {\bibinfo {author} {\bibfnamefont {T.~C.}\ \bibnamefont
  {Lubensky}}\ and\ \bibinfo {author} {\bibfnamefont {J.}~\bibnamefont
  {Isaacson}},\ }\href {\doibase 10.1103/PhysRevA.20.2130} {\bibfield
  {journal} {\bibinfo  {journal} {Phys. Rev. A}\ }\textbf {\bibinfo {volume}
  {20}},\ \bibinfo {pages} {2130} (\bibinfo {year} {1979})}\BibitemShut
  {NoStop}%
\bibitem [{\citenamefont {Redner}(1979)}]{Redner_1979}%
  \BibitemOpen
  \bibfield  {author} {\bibinfo {author} {\bibfnamefont {S.}~\bibnamefont
  {Redner}},\ }\href {\doibase 10.1088/0305-4470/12/9/004} {\bibfield
  {journal} {\bibinfo  {journal} {J. Phys. A}\ }\textbf {\bibinfo {volume}
  {12}},\ \bibinfo {pages} {L239} (\bibinfo {year} {1979})}\BibitemShut
  {NoStop}%
\bibitem [{\citenamefont {Gaunt}(1980)}]{Gaunt_1980}%
  \BibitemOpen
  \bibfield  {author} {\bibinfo {author} {\bibfnamefont {D.~S.}\ \bibnamefont
  {Gaunt}},\ }\href {\doibase 10.1088/0305-4470/13/4/005} {\bibfield  {journal}
  {\bibinfo  {journal} {J. Phys. A}\ }\textbf {\bibinfo {volume} {13}},\
  \bibinfo {pages} {L97} (\bibinfo {year} {1980})}\BibitemShut {NoStop}%
\bibitem [{\citenamefont {{Hsu}}\ \emph {et~al.}(2005)\citenamefont {{Hsu}},
  \citenamefont {{Nadler}},\ and\ \citenamefont
  {{Grassberger}}}]{LatticeAnimals}%
  \BibitemOpen
  \bibfield  {author} {\bibinfo {author} {\bibfnamefont {H.-P.}\ \bibnamefont
  {{Hsu}}}, \bibinfo {author} {\bibfnamefont {W.}~\bibnamefont {{Nadler}}}, \
  and\ \bibinfo {author} {\bibfnamefont {P.}~\bibnamefont {{Grassberger}}},\
  }\href {\doibase 10.1088/0305-4470/38/4/001} {\bibfield  {journal} {\bibinfo
  {journal} {J. of Phys. A}\ }\textbf {\bibinfo {volume} {38}},\ \bibinfo
  {pages} {775} (\bibinfo {year} {2005})},\ \Eprint
  {http://arxiv.org/abs/cond-mat/0408061} {arXiv:cond-mat/0408061
  [cond-mat.stat-mech]} \BibitemShut {NoStop}%
\bibitem [{\citenamefont {{Fytas}}\ and\ \citenamefont
  {{Mart{\'\i}n-Mayor}}(2013)}]{Fytas3}%
  \BibitemOpen
  \bibfield  {author} {\bibinfo {author} {\bibfnamefont {N.~G.}\ \bibnamefont
  {{Fytas}}}\ and\ \bibinfo {author} {\bibfnamefont {V.}~\bibnamefont
  {{Mart{\'\i}n-Mayor}}},\ }\href {\doibase 10.1103/PhysRevLett.110.227201}
  {\bibfield  {journal} {\bibinfo  {journal} {Phys. Rev. Lett.}\ }\textbf
  {\bibinfo {volume} {110}},\ \bibinfo {pages} {227201} (\bibinfo {year}
  {2013})},\ \Eprint {http://arxiv.org/abs/1304.0318} {arXiv:1304.0318
  [cond-mat.dis-nn]} \BibitemShut {NoStop}%
\bibitem [{\citenamefont {Fytas}\ \emph {et~al.}(2016)\citenamefont {Fytas},
  \citenamefont {Martin-Mayor}, \citenamefont {Picco},\ and\ \citenamefont
  {Sourlas}}]{Picco1}%
  \BibitemOpen
  \bibfield  {author} {\bibinfo {author} {\bibfnamefont {N.~G.}\ \bibnamefont
  {Fytas}}, \bibinfo {author} {\bibfnamefont {V.}~\bibnamefont {Martin-Mayor}},
  \bibinfo {author} {\bibfnamefont {M.}~\bibnamefont {Picco}}, \ and\ \bibinfo
  {author} {\bibfnamefont {N.}~\bibnamefont {Sourlas}},\ }\href {\doibase
  10.1103/PhysRevLett.116.227201} {\bibfield  {journal} {\bibinfo  {journal}
  {Phys. Rev. Lett.}\ }\textbf {\bibinfo {volume} {116}},\ \bibinfo {pages}
  {227201} (\bibinfo {year} {2016})},\ \Eprint
  {http://arxiv.org/abs/1605.05072} {arXiv:1605.05072 [cond-mat.dis-nn]}
  \BibitemShut {NoStop}%
\bibitem [{\citenamefont {Fytas}\ \emph {et~al.}(2017)\citenamefont {Fytas},
  \citenamefont {Martin-Mayor}, \citenamefont {Picco},\ and\ \citenamefont
  {Sourlas}}]{Picco2}%
  \BibitemOpen
  \bibfield  {author} {\bibinfo {author} {\bibfnamefont {N.~G.}\ \bibnamefont
  {Fytas}}, \bibinfo {author} {\bibfnamefont {V.}~\bibnamefont {Martin-Mayor}},
  \bibinfo {author} {\bibfnamefont {M.}~\bibnamefont {Picco}}, \ and\ \bibinfo
  {author} {\bibfnamefont {N.}~\bibnamefont {Sourlas}},\ }\href {\doibase
  10.1103/PhysRevE.95.042117} {\bibfield  {journal} {\bibinfo  {journal} {Phys.
  Rev. E}\ }\textbf {\bibinfo {volume} {95}},\ \bibinfo {pages} {042117}
  (\bibinfo {year} {2017})},\ \Eprint {http://arxiv.org/abs/1612.06156}
  {arXiv:1612.06156 [cond-mat.dis-nn]} \BibitemShut {NoStop}%
\bibitem [{\citenamefont {Fytas}\ \emph {et~al.}(2019)\citenamefont {Fytas},
  \citenamefont {Martin-Mayor}, \citenamefont {Parisi}, \citenamefont {Picco},\
  and\ \citenamefont {Sourlas}}]{Picco3}%
  \BibitemOpen
  \bibfield  {author} {\bibinfo {author} {\bibfnamefont {N.~G.}\ \bibnamefont
  {Fytas}}, \bibinfo {author} {\bibfnamefont {V.}~\bibnamefont {Martin-Mayor}},
  \bibinfo {author} {\bibfnamefont {G.}~\bibnamefont {Parisi}}, \bibinfo
  {author} {\bibfnamefont {M.}~\bibnamefont {Picco}}, \ and\ \bibinfo {author}
  {\bibfnamefont {N.}~\bibnamefont {Sourlas}},\ }\href {\doibase
  10.1103/PhysRevLett.122.240603} {\bibfield  {journal} {\bibinfo  {journal}
  {Phys. Rev. Lett.}\ }\textbf {\bibinfo {volume} {122}},\ \bibinfo {pages}
  {240603} (\bibinfo {year} {2019})},\ \Eprint
  {http://arxiv.org/abs/1901.08473} {arXiv:1901.08473 [cond-mat.stat-mech]}
  \BibitemShut {NoStop}%
\bibitem [{\citenamefont {Br{\'e}zin}\ and\ \citenamefont
  {De~Dominicis}(1998)}]{Brezin1998}%
  \BibitemOpen
  \bibfield  {author} {\bibinfo {author} {\bibfnamefont {E.}~\bibnamefont
  {Br{\'e}zin}}\ and\ \bibinfo {author} {\bibfnamefont {C.}~\bibnamefont
  {De~Dominicis}},\ }\href {\doibase 10.1209/epl/i1998-00428-0} {\bibfield
  {journal} {\bibinfo  {journal} {Europhys. Lett.}\ }\textbf {\bibinfo {volume}
  {44}},\ \bibinfo {pages} {13} (\bibinfo {year} {1998})},\ \Eprint
  {http://arxiv.org/abs/cond-mat/9804266} {cond-mat/9804266} \BibitemShut
  {NoStop}%
\bibitem [{\citenamefont {Feldman}(2002)}]{Feldman}%
  \BibitemOpen
  \bibfield  {author} {\bibinfo {author} {\bibfnamefont {D.~E.}\ \bibnamefont
  {Feldman}},\ }\href {\doibase 10.1103/PhysRevLett.88.177202} {\bibfield
  {journal} {\bibinfo  {journal} {Phys. Rev. Lett.}\ }\textbf {\bibinfo
  {volume} {88}},\ \bibinfo {pages} {177202} (\bibinfo {year} {2002})},\
  \Eprint {http://arxiv.org/abs/cond-mat/0010012} {arXiv:cond-mat/0010012
  [cond-mat.dis-nn]} \BibitemShut {NoStop}%
\bibitem [{Note1()}]{Note1}%
  \BibitemOpen
  \bibinfo {note} {We stress that such a SUSY-breaking is explicit and not
  spontaneous.}\BibitemShut {Stop}%
\bibitem [{Note2()}]{Note2}%
  \BibitemOpen
  \bibinfo {note} {Other theoretical ideas and methods used for understanding
  the phase transition in the RF models and the loss of PS SUSY include:
  formation of bound state of replicas \cite {Brezin2001,PSbound}, expansion at
  high temperature \cite {HighT} and around the Bethe lattice \cite
  {Parisi2019}, and the conformal bootstrap \cite
  {Hikami:2017sbg,Hikami:2018mrf}. Comparison to functional renormalization
  group studies will be given at the end.}\BibitemShut {Stop}%
\bibitem [{\citenamefont {Fisher}(1978)}]{Fisher:1978pf}%
  \BibitemOpen
  \bibfield  {author} {\bibinfo {author} {\bibfnamefont {M.}~\bibnamefont
  {Fisher}},\ }\href {\doibase 10.1103/PhysRevLett.40.1610} {\bibfield
  {journal} {\bibinfo  {journal} {Phys. Rev. Lett.}\ }\textbf {\bibinfo
  {volume} {40}},\ \bibinfo {pages} {1610} (\bibinfo {year}
  {1978})}\BibitemShut {NoStop}%
\bibitem [{Note3()}]{Note3}%
  \BibitemOpen
  \bibinfo {note} {A special model of BP with microscopically realized SUSY
  \cite {zbMATH02068689} was proven to undergo dimensional reduction in any
  $d$. This result does not apply to generic (non SUSY) BP models or to the RF
  $\phi ^3$ itself and does not shed light on why PS conjecture works for those
  models.}\BibitemShut {Stop}%
\bibitem [{\citenamefont {Cardy}(1983)}]{CARDY1983470}%
  \BibitemOpen
  \bibfield  {author} {\bibinfo {author} {\bibfnamefont {J.~L.}\ \bibnamefont
  {Cardy}},\ }\href {\doibase https://doi.org/10.1016/0370-2693(83)91328-X}
  {\bibfield  {journal} {\bibinfo  {journal} {Physics Letters B}\ }\textbf
  {\bibinfo {volume} {125}},\ \bibinfo {pages} {470 } (\bibinfo {year}
  {1983})}\BibitemShut {NoStop}%
\bibitem [{\citenamefont {Klein}\ and\ \citenamefont
  {Perez}(1983)}]{KLEIN1983473}%
  \BibitemOpen
  \bibfield  {author} {\bibinfo {author} {\bibfnamefont {A.}~\bibnamefont
  {Klein}}\ and\ \bibinfo {author} {\bibfnamefont {J.~F.}\ \bibnamefont
  {Perez}},\ }\href {\doibase https://doi.org/10.1016/0370-2693(83)91329-1}
  {\bibfield  {journal} {\bibinfo  {journal} {Physics Letters B}\ }\textbf
  {\bibinfo {volume} {125}},\ \bibinfo {pages} {473 } (\bibinfo {year}
  {1983})}\BibitemShut {NoStop}%
\bibitem [{\citenamefont {Klein}\ \emph {et~al.}(1984)\citenamefont {Klein},
  \citenamefont {Landau},\ and\ \citenamefont {Perez}}]{Klein:1984ff}%
  \BibitemOpen
  \bibfield  {author} {\bibinfo {author} {\bibfnamefont {A.}~\bibnamefont
  {Klein}}, \bibinfo {author} {\bibfnamefont {L.~J.}\ \bibnamefont {Landau}}, \
  and\ \bibinfo {author} {\bibfnamefont {J.~F.}\ \bibnamefont {Perez}},\ }\href
  {\doibase 10.1007/BF01403882} {\bibfield  {journal} {\bibinfo  {journal}
  {Commun. Math. Phys.}\ }\textbf {\bibinfo {volume} {94}},\ \bibinfo {pages}
  {459} (\bibinfo {year} {1984})}\BibitemShut {NoStop}%
%%CITATION = CMPHA,94,459;%%
\bibitem [{\citenamefont {Zaboronski}(2002)}]{Zaboronsky:1996qn}%
  \BibitemOpen
  \bibfield  {author} {\bibinfo {author} {\bibfnamefont {O.~V.}\ \bibnamefont
  {Zaboronski}},\ }\href {\doibase 10.1088/0305-4470/35/26/312} {\bibfield
  {journal} {\bibinfo  {journal} {J. of Phys. A}\ }\textbf {\bibinfo {volume}
  {35}},\ \bibinfo {pages} {5511} (\bibinfo {year} {2002})},\ \Eprint
  {http://arxiv.org/abs/hep-th/9611157} {arXiv:hep-th/9611157 [hep-th]}
  \BibitemShut {NoStop}%
\bibitem [{\citenamefont {Kaviraj}\ \emph {et~al.}(2020)\citenamefont
  {Kaviraj}, \citenamefont {Rychkov},\ and\ \citenamefont
  {Trevisani}}]{paper1}%
  \BibitemOpen
  \bibfield  {author} {\bibinfo {author} {\bibfnamefont {A.}~\bibnamefont
  {Kaviraj}}, \bibinfo {author} {\bibfnamefont {S.}~\bibnamefont {Rychkov}}, \
  and\ \bibinfo {author} {\bibfnamefont {E.}~\bibnamefont {Trevisani}},\ }\href
  {\doibase 10.1007/JHEP04(2020)090} {\bibfield  {journal} {\bibinfo  {journal}
  {JHEP}\ }\textbf {\bibinfo {volume} {04}},\ \bibinfo {pages} {090} (\bibinfo
  {year} {2020})},\ \Eprint {http://arxiv.org/abs/1912.01617} {arXiv:1912.01617
  [hep-th]} \BibitemShut {NoStop}%
\bibitem [{\citenamefont {Cardy}(1985)}]{CARDY1985123}%
  \BibitemOpen
  \bibfield  {author} {\bibinfo {author} {\bibfnamefont {J.~L.}\ \bibnamefont
  {Cardy}},\ }\href {\doibase https://doi.org/10.1016/0167-2789(85)90154-X}
  {\bibfield  {journal} {\bibinfo  {journal} {Physica D: Nonlinear Phenomena}\
  }\textbf {\bibinfo {volume} {15}},\ \bibinfo {pages} {123 } (\bibinfo {year}
  {1985})}\BibitemShut {NoStop}%
\bibitem [{Note4()}]{Note4}%
  \BibitemOpen
  \bibinfo {note} {This is clear e.g.~from their propagator mixing different
  powers of momentum, see \cite {Cardy-book}, Eq.~(8.39).}\BibitemShut {Stop}%
\bibitem [{Note5()}]{Note5}%
  \BibitemOpen
  \bibinfo {note} {We note in this respect that replica symmetry breaking is
  proven not to happen in the RFIM \cite {Chatterjee}.}\BibitemShut {Stop}%
\bibitem [{\citenamefont {Bray}\ and\ \citenamefont {Moore}(1985)}]{Bray1985}%
  \BibitemOpen
  \bibfield  {author} {\bibinfo {author} {\bibfnamefont {A.~J.}\ \bibnamefont
  {Bray}}\ and\ \bibinfo {author} {\bibfnamefont {M.~A.}\ \bibnamefont
  {Moore}},\ }\href {\doibase 10.1088/0022-3719/18/28/006} {\bibfield
  {journal} {\bibinfo  {journal} {{J. of Phys. C}}\ }\textbf {\bibinfo {volume}
  {18}},\ \bibinfo {pages} {L927} (\bibinfo {year} {1985})}\BibitemShut
  {NoStop}%
\bibitem [{\citenamefont {Fisher}(1986)}]{Fisher86}%
  \BibitemOpen
  \bibfield  {author} {\bibinfo {author} {\bibfnamefont {D.~S.}\ \bibnamefont
  {Fisher}},\ }\href {\doibase 10.1103/PhysRevLett.56.416} {\bibfield
  {journal} {\bibinfo  {journal} {Phys. Rev. Lett.}\ }\textbf {\bibinfo
  {volume} {56}},\ \bibinfo {pages} {416} (\bibinfo {year} {1986})}\BibitemShut
  {NoStop}%
\bibitem [{\citenamefont {Wilson}\ and\ \citenamefont
  {Kogut}(1974)}]{Wilson:1973jj}%
  \BibitemOpen
  \bibfield  {author} {\bibinfo {author} {\bibfnamefont {K.}~\bibnamefont
  {Wilson}}\ and\ \bibinfo {author} {\bibfnamefont {J.~B.}\ \bibnamefont
  {Kogut}},\ }\href {\doibase 10.1016/0370-1573(74)90023-4} {\bibfield
  {journal} {\bibinfo  {journal} {Phys.Rept.}\ }\textbf {\bibinfo {volume}
  {12}},\ \bibinfo {pages} {75} (\bibinfo {year} {1974})}\BibitemShut {NoStop}%
%%CITATION = PRPLC,12,75;%%
\bibitem [{\citenamefont {Kleinert}\ and\ \citenamefont
  {Schulte-Frohlinde}(2001)}]{Kleinert:2001ax}%
  \BibitemOpen
  \bibfield  {author} {\bibinfo {author} {\bibfnamefont {H.}~\bibnamefont
  {Kleinert}}\ and\ \bibinfo {author} {\bibfnamefont {V.}~\bibnamefont
  {Schulte-Frohlinde}},\ }\href {\doibase 10.1142/4733} {\emph {\bibinfo
  {title} {Critical properties of $\phi^4$ theories}}}\ (\bibinfo  {publisher}
  {World Scientiic},\ \bibinfo {year} {2001})\BibitemShut {NoStop}%
\bibitem [{Note6()}]{Note6}%
  \BibitemOpen
  \bibinfo {note} {Here for simplicity we ignore interaction mixing effects,
  taken into account in the computations described below.}\BibitemShut {Stop}%
\bibitem [{\citenamefont {Kaviraj}\ \emph {et~al.}(2021)\citenamefont
  {Kaviraj}, \citenamefont {Rychkov},\ and\ \citenamefont
  {Trevisani}}]{paper2}%
  \BibitemOpen
  \bibfield  {author} {\bibinfo {author} {\bibfnamefont {A.}~\bibnamefont
  {Kaviraj}}, \bibinfo {author} {\bibfnamefont {S.}~\bibnamefont {Rychkov}}, \
  and\ \bibinfo {author} {\bibfnamefont {E.}~\bibnamefont {Trevisani}},\ }\href
  {\doibase 10.1007/JHEP03(2021)219} {\bibfield  {journal} {\bibinfo  {journal}
  {JHEP}\ }\textbf {\bibinfo {volume} {03}},\ \bibinfo {pages} {219} (\bibinfo
  {year} {2021})},\ \Eprint {http://arxiv.org/abs/2009.10087} {arXiv:2009.10087
  [cond-mat.stat-mech]} \BibitemShut {NoStop}%
\bibitem [{\citenamefont {Kupiainen}\ and\ \citenamefont
  {Niemi}(1983)}]{KUPIAINEN1983380}%
  \BibitemOpen
  \bibfield  {author} {\bibinfo {author} {\bibfnamefont {A.}~\bibnamefont
  {Kupiainen}}\ and\ \bibinfo {author} {\bibfnamefont {A.}~\bibnamefont
  {Niemi}},\ }\href {\doibase https://doi.org/10.1016/0370-2693(83)91527-7}
  {\bibfield  {journal} {\bibinfo  {journal} {Physics Letters B}\ }\textbf
  {\bibinfo {volume} {130}},\ \bibinfo {pages} {380 } (\bibinfo {year}
  {1983})}\BibitemShut {NoStop}%
\bibitem [{app()}]{appendix}%
  \BibitemOpen
  \href@noop {} {\bibinfo  {journal} {{See appendices for more clarification on
  PS SUSY, dimensional reduction, properties of leader operators, leader
  families, and RG computations}}\ }\BibitemShut {NoStop}%
\bibitem [{Note7()}]{Note7}%
  \BibitemOpen
\bibfield  {journal} {  }\bibinfo {note} {This is also supported by
  perturbative \cite {Parisi:1979ka, paper2,paper3} and rigorous Lagrangian
  arguments \cite {KLEIN1983473,Klein:1984ff}.}\BibitemShut {Stop}%
\bibitem [{\citenamefont {Wilson}\ and\ \citenamefont
  {Fisher}(1972)}]{Wilson:1971dc}%
  \BibitemOpen
  \bibfield  {author} {\bibinfo {author} {\bibfnamefont {K.~G.}\ \bibnamefont
  {Wilson}}\ and\ \bibinfo {author} {\bibfnamefont {M.~E.}\ \bibnamefont
  {Fisher}},\ }\href {\doibase 10.1103/PhysRevLett.28.240} {\bibfield
  {journal} {\bibinfo  {journal} {Phys.Rev.Lett.}\ }\textbf {\bibinfo {volume}
  {28}},\ \bibinfo {pages} {240} (\bibinfo {year} {1972})}\BibitemShut
  {NoStop}%
%%CITATION = PRLTA,28,240;%%
\bibitem [{Note8()}]{Note8}%
  \BibitemOpen
  \bibinfo {note} {This was first conjectured in \cite {paper2} and now we have
  a rigorous proof \cite {appendix}.}\BibitemShut {Stop}%
\bibitem [{Note9()}]{Note9}%
  \BibitemOpen
  \bibinfo {note} {\label {T_redundant}The SW leader perturbation originating
  from the superstresstensor, $\protect \mathcal {T}^{\theta {\protect
  \mathaccentV {bar}016{\theta }}}_{\theta {\protect \mathaccentV
  {bar}016{\theta }}}$, deserves a special comment. Naively it has dimension
  $d$ and is marginal. However, it is more properly classified as redundant.
  Its only effect is to rescale factor $H$ in \protect \textup {\hbox
  {\mathsurround \z@ \protect \normalfont (\ignorespaces \ref
  {eq:Cardylag}\unskip \@@italiccorr )}}, which is also a parameter entering
  the superspace metric. This rescaling can be undone by field redefinition and
  has no physical consequences.}\BibitemShut {Stop}%
\bibitem [{\citenamefont {Kaviraj}\ and\ \citenamefont
  {Trevisani}(2022)}]{paper3}%
  \BibitemOpen
  \bibfield  {author} {\bibinfo {author} {\bibfnamefont {A.}~\bibnamefont
  {Kaviraj}}\ and\ \bibinfo {author} {\bibfnamefont {E.}~\bibnamefont
  {Trevisani}},\ }\href@noop {} {\  (\bibinfo {year} {2022})},\ \Eprint
  {http://arxiv.org/abs/2203.12629} {arXiv:2203.12629 [hep-th]} \BibitemShut
  {NoStop}%
\bibitem [{\citenamefont {Kehrein}\ and\ \citenamefont
  {Wegner}(1994)}]{Kehrein:1994ff}%
  \BibitemOpen
  \bibfield  {author} {\bibinfo {author} {\bibfnamefont {S.~K.}\ \bibnamefont
  {Kehrein}}\ and\ \bibinfo {author} {\bibfnamefont {F.}~\bibnamefont
  {Wegner}},\ }\href {\doibase 10.1016/0550-3213(94)90406-5} {\bibfield
  {journal} {\bibinfo  {journal} {Nucl.Phys.}\ }\textbf {\bibinfo {volume}
  {B424}},\ \bibinfo {pages} {521} (\bibinfo {year} {1994})},\ \Eprint
  {http://arxiv.org/abs/hep-th/9405123} {arXiv:hep-th/9405123 [hep-th]}
  \BibitemShut {NoStop}%
%%CITATION = HEP-TH/9405123;%%
\bibitem [{Note10()}]{Note10}%
  \BibitemOpen
  \bibinfo {note} {For $k=2$ and $k$ odd we have $\protect \mathcal {F}_{k} =
  0$, while $\protect \mathcal {F}_{4}\propto \protect \mathcal {N}_4$ has a SN
  leader.}\BibitemShut {Stop}%
\bibitem [{Note11()}]{Note11}%
  \BibitemOpen
  \bibinfo {note} {Loop corrections grow rapidly with the number of fields
  inside the operator, making naive extrapolation to $\epsilon =O(1)$
  questionable. See \cite {Badel:2019oxl} for related recent work.}\BibitemShut
  {Stop}%
\bibitem [{Note12()}]{Note12}%
  \BibitemOpen
  \bibinfo {note} {Actually none of the infinitely many discussed leaders
  become relevant for the $\phi ^3$ case.}\BibitemShut {Stop}%
\bibitem [{Note13()}]{Note13}%
  \BibitemOpen
  \bibinfo {note} {If relevant, this operator would break SUSY since it breaks
  superrotations.}\BibitemShut {Stop}%
\bibitem [{Note14()}]{Note14}%
  \BibitemOpen
  \bibinfo {note} {Feldman \cite {Feldman} argued that SUSY will be lost
  arbitrarily close to $d=6$, because the negative anomalous dimension of
  interactions $\protect \mathcal {F}_k$, $k\ge 8$, grows with $k$ making them
  to cross marginality closer and closer to $d=6$ as $k\to \infty $. We
  disagree with this argument as it does not take into account nonperturbative
  mixing \cite {paper2}. Our new results for the $\protect \mathcal {G}_k$
  family strengthen this objection. The second $\protect \mathbb {Z}_2$
  respecting operator of this family, $(\protect \mathcal {G}_8)_L$, stays
  irrelevant due to its positive one-loop dimension (see \protect \textup
  {\hbox {\mathsurround \z@ \protect \normalfont (\ignorespaces \ref
  {Gk4}\unskip \@@italiccorr )}}). Nonperturbatively (while not in perturbation
  theory) $(\protect \mathcal {G}_8)_L$ is expected to mix with Feldman
  operators, and level crossing will be avoided. Hence, we expect that
  $(\protect \mathcal {G}_8)_L$ will provide a barrier which $(\protect
  \mathcal {F}_k)_L$, $k\ge 8$, cannot cross, remaining irrelevant and
  unimportant for deciding the fate of SUSY.}\BibitemShut {Stop}%
\bibitem [{\citenamefont {Guida}\ and\ \citenamefont
  {Zinn-Justin}(1998)}]{Guida:1998bx}%
  \BibitemOpen
  \bibfield  {author} {\bibinfo {author} {\bibfnamefont {R.}~\bibnamefont
  {Guida}}\ and\ \bibinfo {author} {\bibfnamefont {J.}~\bibnamefont
  {Zinn-Justin}},\ }\href {\doibase 10.1088/0305-4470/31/40/006} {\bibfield
  {journal} {\bibinfo  {journal} {J. Phys. A}\ }\textbf {\bibinfo {volume}
  {31}},\ \bibinfo {pages} {8103} (\bibinfo {year} {1998})},\ \Eprint
  {http://arxiv.org/abs/cond-mat/9803240} {arXiv:cond-mat/9803240} \BibitemShut
  {NoStop}%
\bibitem [{\citenamefont {Kompaniets}\ and\ \citenamefont
  {Panzer}(2017)}]{Kompaniets:2017yct}%
  \BibitemOpen
  \bibfield  {author} {\bibinfo {author} {\bibfnamefont {M.~V.}\ \bibnamefont
  {Kompaniets}}\ and\ \bibinfo {author} {\bibfnamefont {E.}~\bibnamefont
  {Panzer}},\ }\href {\doibase 10.1103/PhysRevD.96.036016} {\bibfield
  {journal} {\bibinfo  {journal} {Phys. Rev. D}\ }\textbf {\bibinfo {volume}
  {96}},\ \bibinfo {pages} {036016} (\bibinfo {year} {2017})},\ \Eprint
  {http://arxiv.org/abs/1705.06483} {arXiv:1705.06483 [hep-th]} \BibitemShut
  {NoStop}%
\bibitem [{\citenamefont {Kompaniets}\ and\ \citenamefont
  {Wiese}(2020)}]{Kompaniets:2019zes}%
  \BibitemOpen
  \bibfield  {author} {\bibinfo {author} {\bibfnamefont {M.}~\bibnamefont
  {Kompaniets}}\ and\ \bibinfo {author} {\bibfnamefont {K.~J.}\ \bibnamefont
  {Wiese}},\ }\href {\doibase 10.1103/PhysRevE.101.012104} {\bibfield
  {journal} {\bibinfo  {journal} {Phys. Rev. E}\ }\textbf {\bibinfo {volume}
  {101}},\ \bibinfo {pages} {012104} (\bibinfo {year} {2020})},\ \Eprint
  {http://arxiv.org/abs/1908.07502} {arXiv:1908.07502 [cond-mat.stat-mech]}
  \BibitemShut {NoStop}%
\bibitem [{\citenamefont {Tissier}\ and\ \citenamefont
  {Tarjus}(2012)}]{TarjusIV}%
  \BibitemOpen
  \bibfield  {author} {\bibinfo {author} {\bibfnamefont {M.}~\bibnamefont
  {Tissier}}\ and\ \bibinfo {author} {\bibfnamefont {G.}~\bibnamefont
  {Tarjus}},\ }\href {\doibase 10.1103/PhysRevB.85.104203} {\bibfield
  {journal} {\bibinfo  {journal} {Phys. Rev. B}\ }\textbf {\bibinfo {volume}
  {85}},\ \bibinfo {pages} {104203} (\bibinfo {year} {2012})},\ \Eprint
  {http://arxiv.org/abs/1110.5500} {arXiv:1110.5500} \BibitemShut {NoStop}%
\bibitem [{\citenamefont {{Balog}}\ \emph {et~al.}(2020)\citenamefont
  {{Balog}}, \citenamefont {{Tarjus}},\ and\ \citenamefont
  {{Tissier}}}]{2020PhRvE.102f2154B}%
  \BibitemOpen
  \bibfield  {author} {\bibinfo {author} {\bibfnamefont {I.}~\bibnamefont
  {{Balog}}}, \bibinfo {author} {\bibfnamefont {G.}~\bibnamefont {{Tarjus}}}, \
  and\ \bibinfo {author} {\bibfnamefont {M.}~\bibnamefont {{Tissier}}},\ }\href
  {\doibase 10.1103/PhysRevE.102.062154} {\bibfield  {journal} {\bibinfo
  {journal} {Phys. Rev. E}\ }\textbf {\bibinfo {volume} {102}},\ \bibinfo
  {pages} {062154} (\bibinfo {year} {2020})},\ \Eprint
  {http://arxiv.org/abs/2008.13650} {arXiv:2008.13650 [cond-mat.dis-nn]}
  \BibitemShut {NoStop}%
\bibitem [{Note15()}]{Note15}%
  \BibitemOpen
  \bibinfo {note} {But not for $d=3$, because in this dimension the SUSY fixed
  point ceases to exist, see \cite {paper2}, Section 3.1.}\BibitemShut {Stop}%
\bibitem [{\citenamefont {Br{\'e}zin}\ and\ \citenamefont
  {De~Dominicis}(2001)}]{Brezin2001}%
  \BibitemOpen
  \bibfield  {author} {\bibinfo {author} {\bibfnamefont {E.}~\bibnamefont
  {Br{\'e}zin}}\ and\ \bibinfo {author} {\bibfnamefont {C.}~\bibnamefont
  {De~Dominicis}},\ }\href {\doibase 10.1007/s100510170323} {\bibfield
  {journal} {\bibinfo  {journal} {Eur. Phys. J. B}\ }\textbf {\bibinfo {volume}
  {19}},\ \bibinfo {pages} {467} (\bibinfo {year} {2001})},\ \Eprint
  {http://arxiv.org/abs/cond-mat/0007457} {cond-mat/0007457} \BibitemShut
  {NoStop}%
\bibitem [{\citenamefont {Parisi}\ and\ \citenamefont
  {Sourlas}(2002)}]{PSbound}%
  \BibitemOpen
  \bibfield  {author} {\bibinfo {author} {\bibfnamefont {G.}~\bibnamefont
  {Parisi}}\ and\ \bibinfo {author} {\bibfnamefont {N.}~\bibnamefont
  {Sourlas}},\ }\href {\doibase 10.1103/PhysRevLett.89.257204} {\bibfield
  {journal} {\bibinfo  {journal} {Phys. Rev. Lett.}\ }\textbf {\bibinfo
  {volume} {89}},\ \bibinfo {pages} {257204} (\bibinfo {year} {2002})},\
  \Eprint {http://arxiv.org/abs/cond-mat/0207415} {arXiv:cond-mat/0207415}
  \BibitemShut {NoStop}%
\bibitem [{\citenamefont {Gofman}\ \emph {et~al.}(1996)\citenamefont {Gofman},
  \citenamefont {Adler}, \citenamefont {Aharony}, \citenamefont {Harris},\ and\
  \citenamefont {Schwartz}}]{HighT}%
  \BibitemOpen
  \bibfield  {author} {\bibinfo {author} {\bibfnamefont {M.}~\bibnamefont
  {Gofman}}, \bibinfo {author} {\bibfnamefont {J.}~\bibnamefont {Adler}},
  \bibinfo {author} {\bibfnamefont {A.}~\bibnamefont {Aharony}}, \bibinfo
  {author} {\bibfnamefont {A.~B.}\ \bibnamefont {Harris}}, \ and\ \bibinfo
  {author} {\bibfnamefont {M.}~\bibnamefont {Schwartz}},\ }\href {\doibase
  10.1103/PhysRevB.53.6362} {\bibfield  {journal} {\bibinfo  {journal} {Phys.
  Rev. B}\ }\textbf {\bibinfo {volume} {53}},\ \bibinfo {pages} {6362}
  (\bibinfo {year} {1996})}\BibitemShut {NoStop}%
\bibitem [{\citenamefont {Angelini}\ \emph {et~al.}(2020)\citenamefont
  {Angelini}, \citenamefont {Lucibello}, \citenamefont {Parisi}, \citenamefont
  {Ricci-Tersenghi},\ and\ \citenamefont {Rizzo}}]{Parisi2019}%
  \BibitemOpen
  \bibfield  {author} {\bibinfo {author} {\bibfnamefont {M.~C.}\ \bibnamefont
  {Angelini}}, \bibinfo {author} {\bibfnamefont {C.}~\bibnamefont {Lucibello}},
  \bibinfo {author} {\bibfnamefont {G.}~\bibnamefont {Parisi}}, \bibinfo
  {author} {\bibfnamefont {F.}~\bibnamefont {Ricci-Tersenghi}}, \ and\ \bibinfo
  {author} {\bibfnamefont {T.}~\bibnamefont {Rizzo}},\ }\href {\doibase
  10.1073/pnas.1909872117} {\bibfield  {journal} {\bibinfo  {journal} {Proc.
  Nat. Acad. Sci.}\ }\textbf {\bibinfo {volume} {117}},\ \bibinfo {pages}
  {2268} (\bibinfo {year} {2020})},\ \Eprint {http://arxiv.org/abs/1906.04437}
  {arXiv:1906.04437 [cond-mat.dis-nn]} \BibitemShut {NoStop}%
\bibitem [{\citenamefont {Hikami}(2018)}]{Hikami:2017sbg}%
  \BibitemOpen
  \bibfield  {author} {\bibinfo {author} {\bibfnamefont {S.}~\bibnamefont
  {Hikami}},\ }\href {\doibase 10.1093/ptep/pty132} {\bibfield  {journal}
  {\bibinfo  {journal} {PTEP}\ }\textbf {\bibinfo {volume} {2018}},\ \bibinfo
  {pages} {123I01} (\bibinfo {year} {2018})},\ \Eprint
  {http://arxiv.org/abs/1708.03072} {arXiv:1708.03072 [hep-th]} \BibitemShut
  {NoStop}%
%%CITATION = ARXIV:1708.03072;%%
\bibitem [{\citenamefont {Hikami}(2019)}]{Hikami:2018mrf}%
  \BibitemOpen
  \bibfield  {author} {\bibinfo {author} {\bibfnamefont {S.}~\bibnamefont
  {Hikami}},\ }\href {\doibase 10.1093/ptep/ptz081} {\bibfield  {journal}
  {\bibinfo  {journal} {PTEP}\ }\textbf {\bibinfo {volume} {2019}},\ \bibinfo
  {pages} {083A03} (\bibinfo {year} {2019})},\ \Eprint
  {http://arxiv.org/abs/1801.09052} {arXiv:1801.09052 [cond-mat.dis-nn]}
  \BibitemShut {NoStop}%
%%CITATION = ARXIV:1801.09052;%%
\bibitem [{\citenamefont {{Brydges}}\ and\ \citenamefont
  {{Imbrie}}(2003)}]{zbMATH02068689}%
  \BibitemOpen
  \bibfield  {author} {\bibinfo {author} {\bibfnamefont {D.~C.}\ \bibnamefont
  {{Brydges}}}\ and\ \bibinfo {author} {\bibfnamefont {J.~Z.}\ \bibnamefont
  {{Imbrie}}},\ }\href@noop {} {\bibfield  {journal} {\bibinfo  {journal}
  {{Ann. Math.}}\ }\textbf {\bibinfo {volume} {158}},\ \bibinfo {pages} {1019}
  (\bibinfo {year} {2003})},\ \Eprint {http://arxiv.org/abs/math-ph/0107005}
  {arXiv:math-ph/0107005} \BibitemShut {NoStop}%
\bibitem [{\citenamefont {Cardy}(1996)}]{Cardy-book}%
  \BibitemOpen
  \bibfield  {author} {\bibinfo {author} {\bibfnamefont {J.~L.}\ \bibnamefont
  {Cardy}},\ }\href@noop {} {\emph {\bibinfo {title} {{Scaling and
  renormalization in statistical physics}}}}\ (\bibinfo  {publisher}
  {Cambridge, UK: Univ. Pr., 238 p.},\ \bibinfo {year} {1996})\BibitemShut
  {NoStop}%
%%CITATION = INSPIRE-429658;%%
\bibitem [{\citenamefont {{Chatterjee}}(2015)}]{Chatterjee}%
  \BibitemOpen
  \bibfield  {author} {\bibinfo {author} {\bibfnamefont {S.}~\bibnamefont
  {{Chatterjee}}},\ }\href {\doibase 10.1007/s00220-014-2269-5} {\bibfield
  {journal} {\bibinfo  {journal} {Comm. Math. Phys.}\ }\textbf {\bibinfo
  {volume} {337}},\ \bibinfo {pages} {93} (\bibinfo {year} {2015})},\ \Eprint
  {http://arxiv.org/abs/1404.7178} {arXiv:1404.7178 [math-ph]} \BibitemShut
  {NoStop}%
\bibitem [{\citenamefont {Badel}\ \emph {et~al.}(2019)\citenamefont {Badel},
  \citenamefont {Cuomo}, \citenamefont {Monin},\ and\ \citenamefont
  {Rattazzi}}]{Badel:2019oxl}%
  \BibitemOpen
  \bibfield  {author} {\bibinfo {author} {\bibfnamefont {G.}~\bibnamefont
  {Badel}}, \bibinfo {author} {\bibfnamefont {G.}~\bibnamefont {Cuomo}},
  \bibinfo {author} {\bibfnamefont {A.}~\bibnamefont {Monin}}, \ and\ \bibinfo
  {author} {\bibfnamefont {R.}~\bibnamefont {Rattazzi}},\ }\href {\doibase
  10.1007/JHEP11(2019)110} {\bibfield  {journal} {\bibinfo  {journal} {JHEP}\
  }\textbf {\bibinfo {volume} {11}},\ \bibinfo {pages} {110} (\bibinfo {year}
  {2019})},\ \Eprint {http://arxiv.org/abs/1909.01269} {arXiv:1909.01269
  [hep-th]} \BibitemShut {NoStop}%
\bibitem [{\citenamefont {Nakayama}(2015)}]{Nakayama:2013is}%
  \BibitemOpen
  \bibfield  {author} {\bibinfo {author} {\bibfnamefont {Y.}~\bibnamefont
  {Nakayama}},\ }\href {\doibase 10.1016/j.physrep.2014.12.003} {\bibfield
  {journal} {\bibinfo  {journal} {Phys. Rept.}\ }\textbf {\bibinfo {volume}
  {569}},\ \bibinfo {pages} {1} (\bibinfo {year} {2015})},\ \Eprint
  {http://arxiv.org/abs/1302.0884} {arXiv:1302.0884 [hep-th]} \BibitemShut
  {NoStop}%
\bibitem [{\citenamefont {Srednicki}(2007)}]{srednicki_2007}%
  \BibitemOpen
  \bibfield  {author} {\bibinfo {author} {\bibfnamefont {M.}~\bibnamefont
  {Srednicki}},\ }\href {\doibase 10.1017/CBO9780511813917} {\emph {\bibinfo
  {title} {Quantum Field Theory}}}\ (\bibinfo  {publisher} {Cambridge
  University Press},\ \bibinfo {year} {2007})\BibitemShut {NoStop}%
\end{thebibliography}%

\vspace{2cm}

%\bibliography{mybib}

\onecolumngrid
\appendix
\clearpage

%\section*{Supplementary material}

\section{PS SUSY and dimensional reduction}

In the main text we had introduced the Parisi-Sourlas (PS) CFT as the IR fixed point of the supersymmetric theory in \eqref{eq:susy}. At the IR fixed point supersymmetry is enhanced to a superconformal symmetry. This enhancement is a supersymmetric counterpart to the familiar emergence of conformal symmetry at the fixed points of non-supersymmetric models (see \cite{Nakayama:2013is} for a review). In the main text we also discussed that there is a dimensional reduction from the PS SUSY CFT to a CFT$_{\widehat{d}}$. In this appendix we clarify how the PS CFT is a simple generalisation of a usual CFT, which allows a straightforward extension of the usual CFT axioms to the SUSY case. Based on that, we provide some details on how a restricted sector of the theory defines a local CFT in $\widehat d$ dimension. The purpose of this appendix, based on \cite{paper1}, is to familiarize the reader with the concept of PS CFT.

\subsection{PS CFT}

Recall that to write the SUSY theory \eqref{eq:susy} in a compact way we had introduced the superspace coordinate  $y^a \equiv (x^\m,\theta,\thetab)$. Here $x^{\mu}\in \mathbb{R}^{d}$ are usual bosonic coordinates while $\theta,\thetab$ are Grassmann-valued (anticommuting) coordinates. The superspace index $a$ takes values $1,\dots d, \theta, \thetab$ while $\m=1,\dots, d$. The OSp$(d|2)$ symmetry preserves the superspace distance $y^ay^b(g_{d|2})_{ab}$. Here the superspace metric is a natural extension of usual flat space metric, given by
\begin{eqnarray}
\label{metricOSp}
&(g_{d|2})_{ab}\equiv 
\left(
\begin{array}{c c}
g_{d} &0 \\
0 &\frac{2}{H}g_{\textrm{Sp}(2)} \\
\end{array}
\right) \ ,  \quad 
%\\
&g_{d} \equiv \mbox{diag}(\overbrace{1,1,\dots, 1}^{d}) \ .
\quad 
g_{\textrm{Sp}(2)} \equiv \left(
\begin{array}{c c}
0 &-1 \\
1 &0 \\
\end{array}
\right) \, .
\end{eqnarray}
%The symmetric scalar product $y_1^a  y_{2 \, a} \equiv y_1 \cdot y_2=y_2 \cdot y_1$ defines a superspace distance that can be expanded in Grassmann variables as $y^2=x^2- 2 \theta \thetab$.
The trace of the metric is computed as $\tr g_{d|2} \equiv (g_{d|2})_a^{\phantom{a}a}=d-2$ (notice that $\tr g_{2|2}=0$).
Derivatives in superspace are defined as $\partial_a \equiv \{\partial_\mu,\partial_\theta,\partial_{\thetab}\} $ 
%\begin{equation}
%\partial_a \equiv  \frac{\partial}{\partial y^a}=\{\partial_\mu,\partial_\theta,\partial_{\thetab}\} \, ,
%\end{equation}
and therefore the super-Laplacian of equation \eqref{eq:susy}  takes the form $\partial^a \partial_a=\partial^\m \partial_\mu-H\partial_\theta\partial_{\thetab}$.

The generators of the supersymmetry are simple extensions of usual Poincar\'e symmetry and they generate supertranslations ($P^a$) and superrotations ($M^{ab}$). Here $P^\mu$ and $M^{\mu\nu}$ are the usual ones, while $P^\theta$, $P^\thetab$, $M^{\mu\theta}$, $M^{\mu\thetab}$, $M^{\theta\thetab}$ are new generators. $P^\theta$, $P^\thetab$ are supertranslaton generators while $M^{\mu\theta}$, $M^{\mu\thetab}$ rotate bosonic into fermionic coordinates and are naturally called superotations.

At the fixed point the theory has the superconformal symmetry of OSp$(d+1,1|2)$. This is again a simple extension of the usual conformal group $SO(d+1,1)$, where we get the extra generators: superdilations $D$ and special superconformal transformations $K^a$, given by:
\be
D=-y^a\partial_a,\quad K^a = 2 y^a y^b\partial_b- y^by_b \partial_a\,.
\ee 
These expressions are thus very similar to the familiar expressions of the usual bosonic conformal symmetry, although note that superconformal $K^\mu$ does not reduce to the bosonic $K_{\rm bos}^\mu = 2 y^\mu y^\nu\partial_\nu- y^\nu y_\nu \partial_\mu$.

%The superspace representations of the generators are straightforward extensions of usual conformal generators - with superspace indices $a,b$ taking the role of Lorentz indices $\mu,\nu$. E.g. one has  $P^a=\partial^a$, $D=-y^a\partial_a$, etc.
The algebra of generators is also similar to that of the conformal algebra $SO(d+1,1)$ but given in terms of graded-commutator $[X,Y\}$ - which is an anticommutator $\{X,Y\}$ if the two generators $X$ and $Y$ involve only the fermionic part of the group, otherwise a commutator $[X,Y]$. E.g. one has $[D,P^a\}=P^a$. The explicit forms of all these generators and their algebra are given in section 3.1 of \cite{paper1}.

In the main text we mentioned that local operators in PS CFT are classified by two labels $\D$ and $\ell$. We already discussed on p.\,\pageref{SW} of the main text how OSp tensors with spin $\ell$ are defined and associated to a Young tableaux. An exhaustive discussion on the structure of OSp tensors can be found in section 3.1 of \cite{paper1}. It is also clear that one can associate to local operators a superconformal dimension $\D$ according to how they transform under superdilations, quite analogous a usual CFT.

In the SUSY CFT one naturally extends the notion of  primaries and descendants. A SUSY operator $\mathcal{O}$  is superprimary if it satisfies $[K^a, \mathcal{O}(0)\}=0$, or a superdescendant  if it is related to a primary by the action of $P^a$. A superprimary and its superdescendants are grouped into a superconformal multiplet. Note that this multiplet may contain more than one operators that are usual primaries, but only one of them is a superprimary and rest are superdescendants.

Superconformal symmetry strongly restricts correlation functions. 
Their functional form matches the one of usual CFT$_d$, provided that points in $\mathbb{R}^d$ are uplifted to superspace $\mathbb{R}^{d|2}$.
E.g. scalar 2-point functions are fixed as  $\la\mathcal{O}_i(y)\mathcal{O}_j(0)\ra=\d_{ij}(y^2)^{-\D}$. Similarly the scalar 3-point functions are fixed up to OPE coefficients $\l_{ijk}$ as 
\be
\la\mathcal{O}_1(y_1)\mathcal{O}_2(y_2)\mathcal{O}_3(y_3)\ra=\l_{123} |y_{12}|^{-\Delta_{1}-\Delta_{2}+\Delta_{3}}|y_{13}|^{-\Delta_{1}-\Delta_{3}+\Delta_{2}}|y_{23}|^{-\Delta_{2}-\Delta_{3}+\Delta_{1}} \, ,
\ee
where $y_{ij}\equiv y_i-y_j$.
The Operator Product Expansion (OPE) is also akin to the usual one e.g. for scalar operators $\Ocal_i$ it schematically reads
\be
\label{superOPE}
\mathcal{O}_1(y)\mathcal{O}_2(0)\sim  \l_{12 \Ocal}  \frac{y_{a_1} \cdots y_{a_{\ell}}\mathcal{O}^{a_1\dots a_\ell}_{\D \ell}(0)}{|y|^{\D_1+\D_2-\D+\ell}} + \text{superdescendants} \, ,
\ee
where we focused on the contribution of a single superprimary and its superdescendants.
Using the OPE, a 4-point function can be expanded in superconformal blocks $G^d_{\D \ell}$ %$g^{d|2}_{\D \ell}$ 
which correspond to the exchange of the supermultiplets of $ \mathcal{O}_{\D \ell}$ in the OPE above. A more general discussion on 2- and 3-point functions of SUSY operators and superconformal blocks can be found in section 3.2 of \cite{paper1}. Our discussion above should hopefully convince the reader that 
PS CFTs satisfy very similar rules compared to the usual CFTs. 

There is however one frequently used rule--unitarity--which does not hold. PS CFTs are necessarily nonunitary, since they violate spin-statistics relation, having anticommuting fields transforming in scalar rather than spinor representations of SO$(d)$. One should not be surprised that the PS CFT is non-unitary, as we have obtained it be taking the zero degree of freedom limit $n\to 0$. Of course unitarity is not a crucial relation from the point of view of statistical physis, and many statistical physics models are known to be non-unitary, PS CFT being just one more example. Being non-unitary, PS CFT is allowed to contain operators with dimensions below unitarity bounds, $\vf$ being prime example, of classical scaling dimension $d/2-2$. 

\subsection{Dimensional reduction}

Below Eq.~\eqref{supfieldexp} of the main text we discussed that a PS SUSY CFT  can be dimensionally reduced to a local CFT$_{\widehat d}$. Here we will explain this procedure in more detail. %indeed well-defined and quite robust.  
Dimensional reduction proceeds as follows. We take any correlator of SUSY operators $\mathcal{O}_i(y_i)$ and restrict $y_i$ to the $(d-2)$-dimensional bosonic subspace $\mathcal{M}_{\widehat d}$ (see the main text). When we do this, the restricted correlator can be interpreted as a correlator of operators $\widehat{\mathcal{O}}_i(\widehat{x}_i)$ of a $\widehat{d}$-dimensional CFT with $\widehat{x}_i\in \mathbb{R}^{\widehat{d}}$:
\be\label{dimred}
\langle\widehat{\mathcal{O}}_1(\widehat{x}_1)\cdots \widehat{\mathcal{O}}_n(\widehat{x}_n)\rangle=\langle \mathcal{O}_1(y_1)\cdots \mathcal{O}_n(y_n) \rangle \big|_{ \mathcal{M}_{\widehat d}} \, .
\ee
%Here $\mathcal{M}_{\widehat d}$, which is defined in the main text, is basically a $\widehat d$-dimensional subspace with bosonic coordinates inside the the PS superspace. 
In the usual CFT context, the procedure of restricting correlators to a subspace is sometimes referred to as `trvial defect', where the word trivial is referred to the fact that we are not introducing any new degrees of freedom living on the defect, unlike for more nontrivial situations such as interfaces, nor are we introducing any nontrivial boundary conditions nor monodromies around the defect. We are just restricting correlators to the subspace. This procedure breaks the symmetry to SO$(d-1,1)\times\text{OSp}(2|2)$. The SO$(d-1,1)$ in the product is recognized as the conformal symmetry of the restricted CFT$_{\widehat d}$, while OSp$(2|2)$ plays the role of a global symmetry.

We will next get rid of the additional OSp$(2|2)$ symmetry. We want to do this for two reasons. First of all we don't expect a generic $\widehat{d}$-dimensional CFT to have such a symmetry. Second, as we will see below, getting rid of this extra symmetry is crucial to ensure that the dimensionally reduced theory is a \emph{local} $\widehat{d}$-dimensional theory. This latter point is nontrivial as trivial defects normally give rise to nonlocal theories, i.e. theories without a local conserved stress tensor.
 
A natural way to accomplish this is to impose the additional requirement that operators $\mathcal{O}_i$ of the PS CFT in the above procedure should be singlets under OSp$(2|2)$. The restricted operators $\widehat{\mathcal{O}}_i$ are then also singlets. We thus got rid of OSp$(2|2)$ symmetry, as it now acts trivially on all kept operators.

To understand this construction consider the example of a rank 2 tensor superprimary $\Ocal_{ab}$. Before restricting it to $\mathcal{M}_{\widehat d}$, we should convert it to an OSp$(2|2)$ singlet. This can be done contracting it with the $\widehat{d}$ dimensional metric $g_{\widehat{d}}^{ab}$, as follows:
\be\label{S0}
\Ocal_{ab} \to g_{\widehat d}^{c a} g_{\widehat d}^{d b} \Ocal_{ab} \,.
\ee
This amounts to setting the $a,b$ indices to $\widehat{d}$-dimensional indices inside $\mathcal{M}_{\widehat d}$.

There are other ways to get OSp$(2|2)$ singlet from $\Ocal_{ab}$, which involve contracting  $\Ocal_{ab}$ or its derivatives with the metric $g_{2|2}^{ab}$, e.g.
\be\label{S1}
g_{2|2}^{ab} \Ocal_{ab}\, , \; \partial^a_{\perp} \partial^b_{\perp}\Ocal_{ab}  \, ,  \;
g_{\widehat d}^{c a}  \partial^b_{\perp}\Ocal_{ab} \, ,  \;
g_{\widehat d}^{c a} g_{\widehat d}^{d b} \partial^2_{\perp}\Ocal_{ab} \, ,  \;
g_{\widehat d}^{c a}  \partial^b_{\perp}(\partial^2_{\perp})^2 \Ocal_{ab} \, ,  \dots\,.
\ee
Here $\partial_{\perp}$ denotes derivatives along directions orthogonal to $\Mcal_{\widehat d}$: $\partial^a_{\perp} = g_{2|2}^{ab} \partial_b$. 

Now, a crucial fact is that any { OSp$(2|2)$ singlet} correlator involving one or more operators of the type \eqref{S1} vanishes when restricted to $\mathcal{M}_{\widehat d}$. This happens because the restriction involves objects like the $g_{2|2}^{ab}$ metric contracted with $\widehat{x}^\mu$ (coordinates  on $\mathcal{M}_{\widehat d}$) or $g_{2|2}^{ab}$ contracted with another $g_{2|2}^{ab}$. All of such contractions are however zero (in particular the supertrace of $g_{2|2}$ is zero). In \cite{paper1} the singlets of the type shown in \eqref{S0} were called $S_0$ operators while the second type i.e. \eqref{S1} as $S_1$.
What we are saying is that if we focus on restricted correlators of $S_0$, all $S_1$ operators decouple.  

To summarize, in dimensional reduction nontrivial operators $\widehat{\mathcal{O}}$ of the reduced theory are $S_0$ singlets obtained from SUSY operators $\Ocal$.  The precise form of the map $\mathcal{O}\to \widehat{\mathcal{O}}$ for a general tensor operator is discussed in section 4.1 of \cite{paper1}.

The decoupling  of $S_1$ operators allows us to  define a stress tensor $\widehat T^{\a\b}$ of the CFT$_{\widehat d}$ from the super-stress tensor  $\mathcal{T}^{ab}$  of the PS CFT. One can write $\widehat T^{\a\b}(\widehat{x})=\mathcal{T}^{\{ \a\b \} }_0(\widehat{x})$  where $\{  \}$ opportunely removes SO$(\widehat d)$-traces, which are $S_1$ operators. 
The super-stress tensor has the superconformal dimension $d-2$ which follows from its superconservation equation  (these properties of $\mathcal{T}$ are discussed in detail in section 3.2 and App. C of  \cite{paper1}). As a consequence
 $\widehat T$ also has  dimension $\widehat d$. Note that it also has SO$(\widehat d)$-spin two. 
 This already indicates that it is a conserved stress stensor in CFT$_{\widehat d}$. To motivate this further,  one can  easily see that it is conserved up to $S_1$ operators, namely $0=\partial_a \mathcal{T}_{0}^{a\b}=\partial_{\a}\widehat{T}^{\a\b}+S_1$. Thus if the PS CFT is local, the reduced theory is also local. For a detailed proof of this conservation equation and a discussion of Ward identities see section 4.3 of \cite{paper1}.

It should be clear by now how a reduced  local CFT is defined from the PS SUSY CFT. To give a more complete picture of the CFT$_{\widehat d}$, we will now show the operator product expansion (OPE) of reduced theory operators. We will see below that this leaves us with some nice consequences.
Let us take the superfield OPE \eqref{superOPE}. We focus on $S_0$ type singlets and restrict the OPE to $\Mcal_{\widehat{d}}$. Then it takes a very simple form:
\begin{equation}
\label{RestrictedOPE}
\!\!\!
\widehat{\mathcal{O}}_1(\widehat{x})\widehat{\mathcal{O}}_2(0) 
\sim \l_{12 \Ocal}   \dfrac{\widehat{x}_{\a_1} \cdots \widehat{x}_{\a_\ell} \widehat{\mathcal{O}}^{\a_1 \dots \a_\ell}_{\D \ell}(0)}{|\widehat{x}|^{\D_1+\D_2-\D+\ell}} + \text{descendants}
+ S_1 \, ,
\end{equation}
where, for each superprimary $\mathcal{O}_{\D \ell}$ exchanged in \eqref{superOPE}, there is in \eqref{RestrictedOPE} a unique operator $\widehat{\mathcal{O}}_{\D \ell}$ of $S_0$ type with the same $\D, \ell$ and $ \l_{12 \Ocal}$ 	as in  \eqref{superOPE}. The other infinitely many primaries are  $S_1$ and thus decouple when the OPE is used in OSp$(2|2)$-invariant corrrelators.
This leads to a remarkable fact: \textit{OSp$(d+1,1|2)$ superconformal blocks are equal to SO$(d-1,1)$ conformal blocks i.e.} $G_{\D \ell}^{d}=g_{\D \ell}^{d-2}$. 
This equality has another beautiful consequence. Note that since a superconformal multiplet packages a finite number of $SO(d+1,1)$ primaries, a superconformal block can be decomposed into a finite number of usual CFT$_{d}$ blocks. We can thus write a CFT$_{d-2}$ block   as a finite combination of  CFT$_{d}$ blocks. This recursion relation is elaborately discussed in section 4.4 of \cite{paper1}.

%%%%%%%%%%%%%%%%%%%%%%
%\section{Leaders and their various properties }\label{sec:leaders}
%\SR{haven't edited this part} In this section we will clarify many of the comments on leader operators that we made in the main text. We will discuss each of the three categories, SW, SN and NSW leaders individually. Our discussion will be based on  classification of the spectrum of all low lying leader operators, up to a certain number of fields and derivatives, that has been carried out in \cite{paper2} and \cite{paper3}. Here we do not repeat the same analysis but pick specific examples to illustrate our points. One important goal in the discussion below is to explain how the families of operators, $(\mathcal{B}_k)_L, (\mathcal{N}_k)_L$ and $(\mathcal{G}_k)_L$ contain the lowest dimensional leaders made of $k$ fields in each category. % respectively and why it was relevant to consider them in the paper.

\section{Susy-writable (SW) leaders}\label{SWapp}

Here we will explain statements about the SW leaders made in the main text and provide some examples. The goal is to make the reader comfortable with this concept.

As defined on p.~\pageref{SW} in the main text, SW leaders are those leader operators which 1) can be mapped to $\psi,\bar{\psi}$ fields and 2) do not vanish after such a map. The first of these conditions means that the operator can only involve $\chi_i$ fields or their derivatives with indices contracted in $O(n-2)$ invariant fashion. E.g. $\chi_i \chi_i$ or $\partial_\mu \chi_i \partial_\mu \chi_i$ can be mapped to $\psi,\bar{\psi}$, becoming, respectively, $2\psi\psibar$ and $2\partial_\mu \psi \partial_\mu \psibar$. An exhaustive discussion of such mapppings, including the origin of the factor 2, is in Appendix C of \cite{paper2}. On the other hand there is no way to map an operator $\sum_{i=2}^{n} \chi_i^3$ to $\psi,\bar{\psi}$ fields. Such a combination is only invariant under S$_{n-1}$ permuting the $\chi_i$'s but not under the $O(n-2)$ rotating them. Leaders containing such combinations of $\chi$'s do exist, and they are classified as non-susy-writable (NSW).

Certain operators can be mapped to $\psi,\psibar$ but vanishes under such a mapping due to the Grassmann nature of the fields. A typical example is $(\chi_i\chi_i)^2$ which maps to zero because $\psi^2=\psibar^2=0$. Leaders having this property are classified not as SW but as susy-null (SN). SN operators have zero correlators among themselves and with any SW operator, but in general not with NSW operators. Hence we cannot just forget about SN leader perturbations, as they might backreact on NSW perturbations and thus, indirectly, destabilize the theory. Indeed, we dedicated a separate section in the main text to the SN leaders (more on this below).

The above definition of SW operators only defines them up to SN contributions. SW leaders with good scaling dimensions (at some order in perturbation theory) will usually be linear combinations of several monomials, some of which can be SN. This subtlety does not play a big role in classifying SW operators and in computing their anomalous dimensions. Under RG flow, SN and SW operators mix triangularly, as SW operators can RG-generate both SW and SN operators, while SN can only generate SN. Thus we can always set the SN part of SW operator to zero in all computations of SW anomalous dimensions. 

Modulo SN operators, the map $\psi \leftrightarrow \chi $ is a bijection in the space of SW operators written in Cardy and in SUSY fields. So we will call SW also the operators in SUSY fields obtained after applying this map.

In the discussion of SW leaders in the main text, a key role was played by the fact that the $\chi  \to  \psi$ map maps them to the highest component of a superfield. This fact was conjectured and extensively checked in \cite{paper2}. A rigorous proof will be presented, for the first time, in App.~\ref{proof} below. Here we will provide some introductory comments, which should make the reader comfortable with this important property.
%We will first discuss  SW leaders and how they can be written as the highest component of SUSY operators after we map to SUSY fields. 
As a first example, consider the S$_n$ singlet operator of the form $\sum_{i=1}^n A(\phi_i)$ which is mapped to Cardy fields as follows
\begin{eqnarray}
 \label{ACardy}
\sum_{i = 1}^n A (\phi_i) & = & A \left( \varphi + \frac{\omega}{2} \right)
+ \sum_{i=2}^n A \left( \varphi - \frac{\omega}{2} + \chi_i \right) 
= \bigg\{A'(\vf) \omega + \frac{1}{2}
A''(\vf)  \sum_{i=2}^n \chi_i^2\bigg\} + \ldots \, .
%\\
%& = & \bigg\{\frac{\delta A}{\delta \varphi} (\varphi) \omega + \frac{1}{2}
%\frac{\delta^2 A}{\delta \varphi^2} (\varphi) \sum_{i=2}^n \chi_i^2\bigg\} + \sum_{k = 3}^{\infty} \frac{1}{k!} \frac{\delta^k A}{\delta
%	\varphi^k} (\varphi) \left[ \left( \frac{\omega}{2} \right)^k + \sum_{i=2}^n \left(
%- \frac{\omega}{2} + \chi_i \right)^k \right] .  \label{ACardy}
\end{eqnarray}
To obtain the r.h.s of this equation we Taylor-expanded the central expression around $\vf =0$. We then gathered terms of the lowest dimension in the curly brackets---this is the leader of the considered S$_n$ singlet. One can easily check that all the terms $\ldots$ have a higher dimension---they are followers.

%All SW leaders are in fact products of such terms. 
Applying the map $ \chi \to \psi $ to the leader, we obtain $A'(\vf)\omega+ A''(\vf)\psi\psib$. Note that this particular combination of fields is invariant under supertranslations transformations: %. In particular it is invariant under the field transformations
\be
\delta\vf= \bar{\varepsilon} \psi+\varepsilon \psib, \qquad \delta\psi=-\varepsilon\omega,\qquad \delta\psib=\bar{\varepsilon}\omega, \qquad  \delta \omega=0 \, .
\ee
These are called supertranslations because they follow from considering how components of the superfield
\be
\Phi(x,\theta,\thetabar)=\vf(x)+\theta \psibar +\thetabar \psi +\theta\thetabar \omega
\ee
transform when applying translations in the fermionic coordinates $\theta,\thetab\to \theta+\varepsilon,\thetab+{\bar{\varepsilon}}$. 

Invariance under supertranslations is a general property of the highest component $\mathcal{O}^{\textbf{a}}_{\theta\thetab}$ of any superfield  $\mathcal{O}^{\textbf{a}}$ (see Eq.~\eqref{supfieldexp}). Since we found that the leader $A'(\vf)\omega+ A''(\vf)\psi\psib$ is supertranslation invariant, it is natural to inquire of which superfield it is the highest component. It is easy to check that the answer is the composite superfield $A(\Phi)$, as stated in the main text.

%This was just one example, but we believe that this must hold generally. Namely any SW leader is supertranslation invariant and hence lives in the highest component of a superfield. In \cite{paper2}, we conjectured this to be true and we tested this extensively on dozens of various SW leader operators considered in that work, including operators containing derivatives. Some more examples are given below. In the future it would be interesting to find a more formal proof of this fact. 

The following additional reasoning may further convince the reader in the plausibility of the discussed property. According to the logic of our approach, leader operators under RG flow generate only leaders. When we specialize to SW leaders, this RG property of the leaders must be encoded in some special selection rule of the SUSY theory. Our claim is that this selection rule is precisely one based on supertranslation invariance.

An additional interesting twist of the story is as follows. The SW leader perturbations we are interested in are scalars under rotation, and they also preserve Sp$(2)$ symmetry rotating $\psi$ and $\psibar$ (this Sp$(2)$ symmetry descends from the $O(n-2)$ symmetry in the $\chi_i$ formulation, after the $n\to0$ limit). In other words, they are SO$(d)\times$Sp$(2)$ invariant fields (although they do not need to respect full OSp$(d|2)$ invariance). As mentioned any such leader lives in the highest component $\mathcal{O}^{\textbf{a}}_{\theta\thetab}$ of a superfield  $\mathcal{O}^{\textbf{a}}$. Note that this superfield does not have to be a scalar. If it is a scalar (no $\textbf{a}$ indices), the leader is just $\mathcal{O}_{\theta\thetab}$. This was the case for the above example. If, on the other hand, the superfield is a tensor, the leader is obtained from $\mathcal{O}^{\textbf{a}}_{\theta\thetab}$ by contracting $\textbf{a}$ indices with the Sp$(2)$ metric, which produces an SO$(d)\times$Sp$(2)$ invariant field as the leader should be. Effectively, to produce a leader we have to set $\textbf{a}$ indices in the directions labelled by $\theta,\thetab$. Hopefully this explains better statements made in this respect in the main text. 
Notice that in principle  the indices $\textbf{a}$ could be contracted also with the SO$(d)$ metric, however this procedure does not generate new operators since $\mathcal{O}$ are supertraceless, which implies $\mathcal{O}^{\mu \mu} = { \frac{4}{H} }%2
\mathcal{O}^{\theta \thetab} $.

Let us consider some examples how this works for tensor superfields. One such superfield, of rank 2, is the superstresstensor $\mathcal{T}^{a b}$ which was defined in App.~C of \cite{paper1}. Computing its highest component $\mathcal{T}_{\theta \thetab}^{a b}$ and setting $a=\theta, b=\thetab$ we obtain
\be
\label{Ttttbtb1}
\mathcal{T}_{\theta \thetab}^{\theta \thetab} \propto \partial_\mu \vf \partial_\mu \omega+   \partial_\mu \psi \partial_\mu \psib+{  \frac{H}{2} (2-d) } \omega^2 %\frac{H}{2}(2-d) \omega^2 \, .
%= \frac{2}{2-d}  [ \partial^\m \vf \partial^\m \omega+   \partial^\m \psi \partial^\m \psib- 4 \omega^2 ] \, ,
\ee
This is clearly a SW leader, since it's a linear combination of terms appearing in the quadratic part of the Parisi-Sourlas SUSY Lagrangian. In fact it is a linear combination of leaders of S$_n$ singlets $\sigma_1^2$ and $\partial^{\m} \phi_i \partial_{\m} \phi_i$.

For a less trivial example, consider the superfield $\Phi^2 \mathcal{T}^{a b}$. In this case the highest component contracted with Sp$(2)$, $(\Phi^2 \mathcal{T}^{\theta \thetab})_{\theta \thetab}$ is a lengthy expression given in Eq.~(8.13) of \cite{paper2}. It can be shown to be a linear combination of two dimension 8 leaders associated with $\sigma_1\sigma_3$ and $\sum_i \phi_i^2\partial_\mu \phi_i \partial_\mu \phi_i$.

Our final example is the Sp$(2)$-invariant part $\Bcal^{\theta \thetab, \theta \thetab}_{\theta \thetab}$ of the highest component $\Bcal^{a b, c d}_{\theta \thetab}$ of the superfield $\Bcal^{a b, c d}$ in the box $(2,2)$ representation. An infinite family of such composite superfields can be built from the fundamental superfield $\Phi$ by an immediate counterpart of the $\widehat{d}$-dimensional Eq.~\eqref{Box-n} in the main text:
\begin{equation}
 (\Bcal^{(k)})^{a b , c  d} \equiv {\Phi}^{k-3} \left( \Phi_{, a b} \Phi_{,
		cd} \Phi - \tfrac{2 \widehat{d}}{\widehat{d} - 2} \Phi_{, a}
	\Phi_{, b} \Phi_{, c d}  \right)_Y \,.
	\label{Box-n1}
\end{equation}
From this equation, one can work out $(\Bcal^{(k)})^{\theta \thetab, \theta \thetab}_{\theta \thetab}$ explicitly in terms of $\vf$, $\psi$, $\psibar$ and $\omega$. For example, the full expression for $k=4$ is given in Eq.~(H.5) of \cite{paper2} and we do not copy it here. It is then possible to check that this expression is (up to total derivatives) a linear combination of SW leaders of the following S$_n$ singlet operators: $\sigma_1^2 \sigma_2$, $ \partial_\mu \sigma_1 \partial_\mu \sigma_3$, $\sigma_1 \sum_i \phi_i \partial_\mu \phi_i \partial_\mu \phi_i$, $\sigma_2 \sum_i \partial_\mu \phi_i \partial_\mu \phi_i$, $\sum_i \phi_i \partial_\mu \phi_i \partial_\nu \phi_i \partial_\mu\partial_\nu \phi_i$, $\sum_i  \phi^2_i \partial_\mu\partial_\nu \phi_i \partial_\mu\partial_\nu \phi_i$. Leader tables from App.~D of \cite{paper2} are helpful when performing this check.

To summarize, in this appendix we explained in more detail how SW leaders are captured by the highest components of superfields (see App.~\ref{proof} for a rigorous proof). Since the superfield scaling dimensions can be found from the dimensionally reduced theory, this property drastically simplifies the task of computing SW leaders scaling dimensions. This is how it was used in the main text.

\subsection{Correspondence between SW leaders and supertranslation invariant SUSY fields}
\label{proof}

For a SW leader $\mathcal{O}$, written in terms of Cardy fields, denote by
$\mathcal{O}' $ the corresponding operator mapped to SUSY fields by the
$\chi \rightarrow \psi$ map. The operation $\mathcal{O} \rightarrow
\mathcal{O}'$ has two properties: 1) the resulting operator $\mathcal{O} '$ is
a supertranslation-invariant (st-invariant, for short) SUSY field; 2) any
st-invariant and Sp(2) invariant field of the SUSY theory can be written as
$\mathcal{O}'$ for some SW leader $\mathcal{O}$. These facts were first
noticed in {\cite{paper2}} based on many examples, and conjectured to be
always true. Here we will provide a rigorous proof that this is indeed the
case.

\subsubsection{Any SW leader maps to an st-invariant operator}

The most general singlet interaction, in terms of replicated fields $\phi_i$,
can be written as a linear combination of products of elementary singlet
interactions, of the form
\begin{equation}
	\sum_{i = 1}^n A [\phi_i] \label{elsingl},
\end{equation}
where $A [\phi ]$ is a function of $\phi$ and its first, second,{\ldots}
derivatives, evaluated at point $x$. Some of the derivative indices may be
contracted with each other, others when products of elementary interactions
are taken. E.g. we may have $A [\phi] = \phi \partial_{\mu} \phi$.

Let us translate \eqref{elsingl} to the Cardy fields. Generalizing Eq.
(5.14) in {\cite{paper2}} to the case when $A [\phi]$ may also depend on the
derivatives, the lowest dimension part of the above singlet is given by
\begin{equation}
	\int \frac{\delta A [\varphi] (x)}{\delta \varphi (x_1)} \omega (x_1) d x_1
	+ \frac{1}{2} \int \frac{\delta^2 A [\varphi] (x)}{\delta \varphi (x_1)
		\delta \varphi (x_2)} \chi_i (x_1) \chi_i (x_2) d x_1 d x_2 \,.
\end{equation}
The $\chi \rightarrow \psi$ map maps this to (use App. C of {\cite{paper2}},
first Eq. (C.3), simplifying as $\frac{\delta^2 A (x)}{\delta \varphi (x_1)
	\delta \varphi (x_2)}$ is symmetric in $x_1, x_2$):
\begin{equation}
	\mathcal{O}' = \int \frac{\delta A (x)}{\delta \varphi (x_1)} \omega (x_1) d
	x_1 + \int \frac{\delta^2 A (x)}{\delta \varphi (x_1) \delta \varphi (x_2)}
	\psi (x_1) \bar{\psi} (x_2) d x_1 d x_2\,.
\end{equation}
Let us check that this is invariant under the supertranslations
\begin{equation}
	\delta_{\tmop{st}} \varphi = \bar{\varepsilon} \psi + \varepsilon
	\overline{\psi,} \quad \delta_{\tmop{st}} \bar{\psi} = \bar{\varepsilon}
	\omega, \quad \delta_{\tmop{st}} \psi = - \varepsilon \omega, \quad
	\delta_{\tmop{st}} \omega = 0 \, .
\end{equation}
Using $\delta_{\tmop{st}} \frac{\delta A (x)}{\delta \varphi (x_1)} = \int
\frac{\delta^2 A (x)}{\delta \varphi (x_1) \delta \varphi (x_2)}
\delta_{\tmop{st}} \varphi (x_2) d x_2$ we get that $\delta_{\tmop{st}}$ of
the first term in $\mathcal{O}'$ cancels with
\begin{equation}
	\int \frac{\delta^2 A (x)}{\delta \varphi (x_1) \delta \varphi (x_2)}
	\delta_{\tmop{st}} [\psi (x_1) \bar{\psi} (x_2)] d x_1 d x_2 .
\end{equation}
Let us show that the remaining part of $\delta_{\tmop{st}}$ of the second term
in $\mathcal{O}'$,
\begin{equation}
	\int \delta_{\tmop{st}} \frac{\delta^2 A (x)}{\delta \varphi (x_1) \delta
		\varphi (x_2)} \psi (x_1) \bar{\psi} (x_2) d x_1 d x_2 = \int \frac{\delta^3
		A (x)}{\delta \varphi (x_1) \delta \varphi (x_2) \delta \varphi (x_3)} \psi
	(x_1) \bar{\psi} (x_2) [\bar{\varepsilon} \psi (x_3) + \varepsilon
	\bar{\psi} (x_3)] \label{d3A},
\end{equation}
vanishes. If $A$ is a function of only $\varphi$ and not of derivatives, then
$\frac{\delta^3 A (x)}{\delta \varphi (x_1) \delta \varphi (x_2) \delta
	\varphi (x_3)}$ is proportional to delta functions, and \eqref{d3A} vanishes
because of $\psi^2 = 0$, $\bar{\psi}^2 = 0$. When $A$ also depends on
derivatives, $\frac{\delta^3 A (x)}{\delta \varphi (x_1) \delta \varphi (x_2)
	\delta \varphi (x_3)}$ will involve derivative of deltas. The important thing
is that it is always symmetric in the interchanges of $x_i$. The first
fermionic product in \eqref{d3A} is antisymmetric in $x_1, x_3$ and the second
in $x_2, x_3$. So the integral will vanish.

The above argument shows that $\mathcal{O}'$ is st-invariant. Products of
$\mathcal{O}'$'s will also be st-invariant. This proves st-invariance of all
SW leaders of general singlets, since those can be obtained by taking linear
combinations of products of different elementary singlet interactions of the
form \eqref{elsingl}.

It may sometimes happen that two different singlet interactions give rise,
through the above construction, to leaders which are either 1) exactly the
same when written in the $\chi$ fields, or 2) becomes the same after $\chi
\rightarrow \psi$. Then, taking their difference, we obtain an interaction
whose leader is either susy-null (case 2), or non-susy-writable (case 1). E.g.
in case 1 the leader will involve contributions of terms $\sum \chi_i^k$, $k
\geqslant 3$, or their generalizations with derivatives, see {\cite{paper2}},
last line of (5.14). These terms cannot cancel because operators $\sum
\chi_i^k$ are all algebraically independent. This shows that the leader will
be NSW. So we need not be worried about such cancelations for the purposes of
proving the result of this section.

\subsubsection{Any Sp(2) and st-invariant operator comes from a SW leader}

To prove this result, we take an arbitrary Sp(2) and st-invariant SUSY
operator. We can write it as $(X)_{\theta \bar{\theta}}$, the $\theta
\bar{\theta}$ component of a Sp(2) invariant superfield $X$. Being Sp(2)
invariant, $X$ can be written as a linear combination of products of the
fundamental superfield $\Phi$ and of its superderivatives, contracted with
either SO$(d)$ metric $g_{\mu \nu}$ or with the Sp(2) metric $g_{a b}$. In
what follows $a, b$ are Sp(2) indices: $a, b \in\{ \theta, \bar{\theta}\}$.

Consider first the simplest case when $X$ only involves normal derivatives,
contracted with $g_{\mu \nu}$. We can simplify this even further by
considering a product of $K$ superfields at different points. We will call such special $X$'s by a letter $Y$:
\begin{equation}
	Y = \Phi (x_1) \ldots \Phi (x_K) \, . \label{Xsimple}
\end{equation}
We will prove that $Y_{\theta \bar{\theta}}$ arises from an SW leader of a
singlet. Differentiating with respect to $x_k$, and setting all ${x_k} $
equal, we can then obtain the statement for $X$'s involving derivatives in the $\mu$ directions.

Let us compute $Y_{\theta \bar{\theta}}$ for \eqref{Xsimple}. We have
\begin{eqnarray}
	Y_{\theta \bar{\theta}} & = & \sum_k \Phi_{\theta \bar{\theta}} (x_k)
	\prod_{l \neq k} \Phi_0 (x_l) + \sum_{k_1 \neq k_2} \Phi_{\theta} (x_{k_1})
	\Phi_{\bar{\theta}} (x_{k_2}) \prod_{l \neq k_1 k_2} \Phi_0 (x_l)
	\nonumber\\
	& = & \sum_k \omega (x_k) \prod_{l \neq k} \varphi (x_l) + \sum_{k_1 \neq
		k_2} \psi (x_{k_1}) \bar{\psi} (x_{k_2}) \prod_{l \neq k_1 k_2} \varphi
	(x_l) \label{susylead}  \,. 
\end{eqnarray}
Consider the following S$_n$ singlet:
\begin{equation}
	\sum_i \phi_i (x_1) \ldots \phi_i (x_K) \, .
\end{equation}
Mapping it to Cardy fields, we obtain the leader
\begin{equation}
	\sum_k \omega (x_k) \prod_{l \neq k} \varphi (x_l) + \sum_{k_1 \neq k_2}
	\sum_i \frac{1}{2} \chi^i (x_{k_1}) \chi^i (x_{k_2}) \prod_{l \neq k_1 k_2}
	\varphi (x_l) \, , \label{lead1}
\end{equation}
plus followers involving terms higher order in $\chi$. According to the
dictionary of App. C of {\cite{paper2}}, the $\chi \rightarrow \psi$ map maps
\begin{equation}
	\chi^i (x_{k_1}) \chi^i (x_{k_2}) \rightarrow \psi (x_{k_1}) \bar{\psi}
	(x_{k_2}) + \psi (x_{k_2}) \bar{\psi} (x_{k_1}) \, .
\end{equation}
Thus we see that \eqref{lead1} maps precisely on \eqref{susylead}. This proves
that any $Y_{\theta \bar{\theta}}$ arises
from an SW leader of a singlet. As mentioned above, using differentiation
w.r.t. $x_k$ we can then prove this statement for $X_{\theta \bar{\theta}}$
arising from any $X$ of the form ${D } ^{(1)} \Phi (x) \ldots D ^{(K)} \Phi (x)$
where $D^{(k)}$ are arbitrary differential operators in the $x$ direction.

Let us now consider $X$ involving some derivatives in the Sp(2) directions. We
will use the same trick as above, eliminating all $x$ derivatives by
considering superfields at $K$ separated points. Thus we are reduced to considering
\begin{equation}
	X = R  ^{(1)} \Phi (x_1) \ldots R ^{(K)} \Phi (x_K) \label{Xwithader} \, ,
\end{equation}
where $R^{(k)}$ are arbitrary differential operators in the $\theta,
\bar{\theta}$ directions, with indices contracted in an Sp(2) invariant way.

We will show the following lemma: \emph{Any ${X_{\theta
		\bar{\theta}}} $ with $X$ of the form \eqref{Xwithader} can be written as a linear
combination of products:
\begin{equation}
	(Y_1)_{\theta \bar{\theta}} \ldots (Y_L)_{\theta \bar{\theta}}\,, \label{products}
\end{equation}
where $Y_l$ are as in \eqref{Xsimple} i.e.~they do not contain any derivatives in the $\theta, \bar{\theta}$
direction (nor in $x$ direction since we are considering separated points).} This statement about the SUSY theory is interesting in its own right, as it means that any Sp(2) and st-invariant field can be written as a linear combination of products of elementary such fields, at most bilinear in $\psi$ and $\psib$. It also allows us to finish the proof. Indeed, above we have shown that any such $(Y_l)_{\theta \bar{\theta}}$ can be written as a SW leader of an S$_n$ singlet interactions. Taking the linear combination of products of
these interactions, we then represent ${X_{\theta \bar{\theta}}}$ as a SW leader. 

So it remains to show the lemma. Note that all derivatives of the order higher than second are zero. Consider next an $X$ containing some second derivatives. We have
\begin{equation}
	\partial_a \partial_b \Phi (x ) = g^{\tmop{Sp} (2)}_{a b} \omega (x) ,\quad a,b\in\{\theta,\thetab\}\,.
\end{equation}
Using this fact, any $X$ with second derivatives can be written as a product of a bunch of $\omega (x_k)
\equiv \Phi (x_k)_{\theta \bar{\theta}}$ times $X'$ which only involves up to first derivatives contracted in Sp(2) invariant fashion. We now have
\begin{equation}
	X_{\theta\thetab}=( X' \prod \Phi (x_k)_{\theta \bar{\theta}} ) _{\theta\thetab} = 
		( X')_{\theta\thetab} \prod \Phi (x_k)_{\theta \bar{\theta} }\,.
\end{equation}

The previous equation reduced the lemma to the case of $X$'s having only up to first derivatives.
Let $X$ be such field. For a recursive step we factor out one product $g^{a b} \partial_a \Phi (x_{k_1}) \partial_b \Phi (x_{k_2})$ from $X$. We renumber points so that $k_1=1$, $k_2=2$ and write
\begin{equation}
	X = g^{{\tmop{Sp} (2)}}_{a b} \partial^a \Phi_1 \partial^b \Phi_2\times X' \, ,
\end{equation}
where $X'$ contains two derivatives less than $X$, and we denoted $\Phi_k \equiv \Phi (x_k)$. We have %Up to an $H$ factor, we have
\begin{equation}
	X =
	(\partial_{\theta} \Phi_1 \partial_{\bar{\theta}} \Phi_2 + \partial_{\theta}
	\Phi_2 \partial_{\bar{\theta}} \Phi_1) X' \, .
\end{equation}
From this expression we see that $X$ is symmetric in $x_1, x_2$.
Let us now express $(X)_{\theta \bar{\theta}}$ as follows
\begin{equation}
	X_{\theta \bar{\theta}} = (\partial ^a \Phi_1 \partial_a \Phi_2 X')_{\theta
		\bar{\theta}} = (\partial ^a [\Phi_1 \partial_a \Phi_2 X'])_{\theta
		\bar{\theta}} - (\Phi_1 \partial^a \partial_a \Phi_2 X')_{\theta
		\bar{\theta}} + (\Phi_1 \partial_a \Phi_2 \partial^a X')_{\theta
		\bar{\theta}} \, .
\end{equation}
The first term vanishes since $()_{\theta \bar{\theta}} =
\partial_{\bar{\theta}} \partial_{\theta} ()$. We also have $(\Phi_1
\partial^a \partial_a \Phi_2 X')_{\theta \bar{\theta}} = (\Phi_2)_{\theta
	\bar{\theta}} (\Phi_1 X')_{\theta \bar{\theta}}$. Symmetrizing in $x_1, x_2$
which is allowed since we know it's symmetric, we have
\begin{eqnarray}
	2 X_{\theta \bar{\theta}} & = & - (\Phi_2)_{\theta \bar{\theta}} (\Phi_1
	X')_{\theta \bar{\theta}} - (\Phi_1)_{\theta \bar{\theta}} (\Phi_2 X')_{\theta
		\bar{\theta}} + (\Phi_1 \partial_a \Phi_2 \partial^a X')_{\theta
		\bar{\theta}} + (\Phi_2 \partial_a \Phi_1 \partial^a X')_{\theta
		\bar{\theta}} \nonumber\\
	& = & - (\Phi_2)_{\theta \bar{\theta}} (\Phi_1 X')_{\theta \bar{\theta}} -
	(\Phi_1)_{\theta \bar{\theta}} (\Phi_2 X')_{\theta \bar{\theta}} - (\Phi_1
	\Phi_2 \partial_a \partial^a X')_{\theta \bar{\theta}} + (\partial_a [\Phi_1
	\Phi_2 \partial^a X'])_{\theta \bar{\theta}} \nonumber\\
	& = & - (\Phi_2)_{\theta \bar{\theta}} (\Phi_1 X')_{\theta \bar{\theta}} -
	(\Phi_1)_{\theta \bar{\theta}} (\Phi_2 X')_{\theta \bar{\theta}} + (\Phi_1
	\Phi_2)_{\theta \bar{\theta}} (X')_{\theta \bar{\theta}} \, ,  \label{master}
\end{eqnarray}
where we took into account that $(\partial_a [\Phi_1 \Phi_2 \partial^a
X'])_{\theta \bar{\theta}} = 0$.

Eq.~\eqref{master} accomplishes a recursive step, since every field in the r.h.s.~contains fewer derivatives than $X$. Applying this formula recursively, we can eliminate all first derivatives. This proves the lemma, and with this the fact in the title of the section.

In conclusion we would like to indicate
that there is an alternative way to find an S$_n$ singlet interaction of which
$X_{\theta \bar{\theta}}$ is the leader, not relying on the above lemma. For this we take $X$ and write it in
Cardy fields. The resulting expression is SO$(n - 1)$ invariant but not, in
general, S$_n$ invariant since it does not in general respect $\phi_1
\leftrightarrow \phi_i$ permutations $P_i$, $i = 2 \ldots n$. These
permutations in Cardy fields are given in Eq.~\eqref{n=0sym} (for $i = 2$). The idea then is to consider the symmetrized linear
combination:
\begin{equation}
	\tilde{X} = X + \sum_{i = 2}^n P_i X,
\end{equation}
which is fully S$_n$ invariant by construction. It can be shown the leader of
$\tilde{X}$ coincides with $X_{\theta \bar{\theta}}$. We omit the details.

\section{Susy-null (SN) and non-susy-writable (NSW) leaders}

We will now focus our attention on SN and NSW leaders. We will highlight some aspects of the SN leaders that may not have been clear from the main text. We will also clarify why we focused on some specific families of  SN and NSW leaders in the main text. An exhaustive version of our discussion below can be found in the appendices of \cite{paper2} and in a subsequent paper \cite{paper3}.

Let us first consider general operators that are SN. We have said on p.\,\pageref{susynull} that, similar to SW operators, SN operators involve contractions of $\chi_i$ fields in an $O(n-2)$ invariant way which allows us to use the map $\chi\to \psi$ to SUSY fields. However SN operators vanish under this map. The simplest example of an SN operator is $(\chi_i\chi_i)^2$ that maps to $(2\psi\psib)^2$ which is zero as $\psi$ and $\psib$ are anticommuting. SN operators may also contain derivatives, e.g. the operator $(\partial_\mu\chi_i\partial^\mu\chi_i)^2$ is SN as it maps to $(\partial_\mu\psi\partial^\mu\psib)^2$ which is zero.

Previously in appendix \ref{SWapp} we commented that SW operators are defined only up to SN operators.  To see what this means consider two different quartic operators: $(\chi_i\partial_{\mu}\chi_i)(\chi_j\partial^{\mu}\chi_j)$ and $(\chi_i\chi_i)(\partial_{\mu}\chi_j\partial^{\mu}\chi_j)$, summation over $i,j$ from 2 to $n$ being understood (note that these operators are not leaders, but this is unimportant for the point we are trying to make). These are clearly SW as $\chi_i\partial_{\mu}\chi_i$ ($=\frac 12\partial_\mu(\chi_i\chi_i)$) and $\partial_{\mu}\chi_j\partial^{\mu}\chi_j$ are SW. Under the $\chi\to \psi$ map it is easy to see that both quartic operators become $\propto \psi\psib\partial_\mu\psi\partial^\mu\psib$. Since they were distinct operators in $\chi$ fields their difference must be SN. This SN operator is in fact $\partial^2(\chi_i\chi_i)^2$. This happens to be a leader of the singlet operator $\partial^2\mathcal{N}_4$ (we defined  $\mathcal{N}_k$ in \eqref{def:NkSingl}). As it is a total derivative we can ignore it as a perturbation in the RG flow.

As for SN \emph{leaders}, in the main text we focused primarily on the family $(\mathcal{N}_k)_L=(\chi_i^2)^2 \vf^{k-4}$. This was because they are the lowest dimensional SN leaders made of $k$ fields. 
This is easy to see since they are built as a product of the lowest dimensional SN operator $(\chi_i^2)^2$ times powers of the lowest dimensional field, thus any lower dimensional operator cannot be SN.

However it is possible to have other SN leaders with a higher classical dimension than $(\mathcal{N}_k)_L$ for the same number of fields. E.g. at the level of $k=5$ fields one can have $\omega(\chi_i\chi_i)^2$ which is a leader of the $\mathbb{Z}_2$ odd singlet $4\s_3\s_1^2-3\s_2^2\s_1$\,. Also with $k=6$ fields one can have $\vf\omega(\chi_i\chi_i)^2, (\chi_i\chi_i)^3$ and $(\partial_\mu\vf)^2(\chi_i\chi_i)^2$. They all have the same classical dimension and hence mix perturbatively. Their singlets are shown explicitly in App. D of \cite{paper2}. We have coumputed the anomalous dimensions of many such operators and checked that they are all positive - so they do not affect our conclusion in the main text.

Passing now to the NSW leaders, one of the families we focused on was $(\mathcal{G}_k)_L\propto \vf^{k-6} (\mathcal{F}_6)_L$. Let us explain the claim from the main text that these are the lowest dimensional NSW leaders made of $k$ fields. This is not immediately obvious since there exist lower dimensional NSW \emph{operators} made of $k$ fields. However it turns out that these operators  (e.g. $\chi_i^3 \vf^{k-3}$) are not leaders.
This can be proven using S$_n$ symmetry. We consider the action of S$_n$ replica symmetry in Cardy fields.
Permutations $\phi_i \leftrightarrow \phi_j $ for $i,j>1$ act as  $\chi_i \leftrightarrow \chi_j$ which mean that we always need $\chi_i$ to appeared in a permutation-symmetric way.
The permutations $\phi_1 \leftrightarrow \phi_j$ acts as (we focus on $j=2$ for convenience), see Eq. (5.3) in \cite{paper2},
\begin{equation}
\varphi \to \varphi + (\chi_2-\omega )  \, , 
\qquad
\omega \to \omega \, ,
\qquad
\chi_2 \to 2 \omega - \chi_2 \, , 
\qquad
\chi_i \to \chi_i - \chi_2 + \omega  \, , 
\label{n=0sym}
\end{equation}
where $i = 3 \ldots n$ and where we set $n=0$. This is less transparent in Cardy fields, nevertheless all S$_n$ singlets at $n=0$ must be invariant under \eqref{n=0sym}.
By  keeping only the lowest dimensional term of \eqref{n=0sym} we obtain a simpler transformation 
\begin{equation}
\varphi \to \varphi , 
\qquad
\omega \to \omega,
\qquad
\chi_2 \to  - \chi_2, 
\qquad
\chi_i \to \chi_i - \chi_2 \, .
\label{n=0symLEAD}
\end{equation}
A necessary condition for an operator to be a leader is that it must be invariant under \eqref{n=0symLEAD}. 
This map acts trivially on all SW (and thus also SN) operators: indeed $\vf,\omega$ and $\chi_i^2$ are left invariant by  \eqref{n=0symLEAD}.
However it can be used to rule out some NSW leader candidates: operators of the form $\Ocal \vf^a \omega^b (\chi_i^2)^c$ cannot be leaders if operator $\Ocal$ is not invariant under \eqref{n=0symLEAD}. 
We thus find that $\chi_i^3 \vf^{k-3}$ are not leader operators because $\chi_i^3$ is not invariant under   \eqref{n=0symLEAD}.
Similarly we can rule out all possible NSW combinations of less than six $\chi_i$ fields since they are not invariant under   \eqref{n=0symLEAD}.
One thus easily recovers that  $(\mathcal{G}_k)_L$ are the lowest dimensional NSW leaders made of $k$ fields.

%E.g. since $\chi_i^3$ is not invariant under   \eqref{n=0symLEAD} we automatically know that $\chi_i^3 \vf^a \omega^b (\chi_i^2)^c$ are not leader operators.
%This observation can be used to prove that the lowest dimensional NSW leaders made of $k$ fields must be the $(\mathcal{G}_k)_L \propto \vf^{k-6} (\Fcal_6)_L$. 
%To build a lower dimensional operator with the same number of fields one would have to trade some $\chi_i$ fields by some $\vf$ fields but the resulting operators cannot be NSW leaders since there is no NSW operator with less than six $\chi_i$ fields which is  invariant under   \eqref{n=0symLEAD}.

The other NSW leader family considered in the main text was $(\mathcal{F}_k)_L$. This one also has an important role: these operators are built out of special linear combinations of $\chi_i$ which very non-trivially are invariant under \eqref{n=0symLEAD}. Indeed we checked up to $k=20$ that $(\mathcal{F}_k)_L$ are the only operators of the form  $\sum_{l=0}^k c_{l} (\chi_i^{k-l}) (\chi_j^l)$ which are invariant under \eqref{n=0symLEAD}, and thus they are the only leaders of this form.  
Of course any operator built by taking powers of $\vf,\omega, \chi_i^2$ multiplied with any $(\mathcal{F}_k)_L$ to any power will respect the invariance under \eqref{n=0symLEAD}. The resulting operators are thus possible leaders of the theory.
It is however important to stress that the invariance under \eqref{n=0symLEAD} is only a necessary condition and that in order to get a good leader operator, one must be able to write it in an S$_n$ invariant way which is thus invariant under the full transformation \eqref{n=0sym}.  E.g. while $\vf \omega$ and $\chi_i^2$ are both  invariant under \eqref{n=0symLEAD}, only the combination $2 \vf \omega+  \chi_i^2$ is a good S$_n$ singlet invariant under  \eqref{n=0sym}.

\section{RG computations}\label{sec:intro}

On p.\,\pageref{Nk3} of the main text we presented a number of results on one- or two-loop corrections to the dimensions of SN and NSW operators. All those results were obtained from perturbative RG computations using standard Feynman diagrammatic approach in dimensional regularization. In this appendix we give a flavor of these computations by discussing some fundamental tools and a few examples. The goal is to show that our results are indeed quite straightforward to obtain. 

\begin{figure}[h]
	\centering \includegraphics[width=300pt]{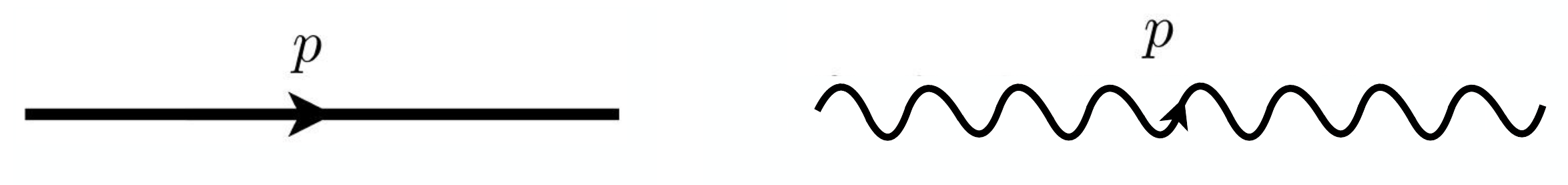}
	\
	\caption{Propagators $G_{\vf\vf}(p)$ (left) and $G_{\chi\chi}(p)$ (right).\label{propagators}
		\label{Fig:propagators}
	}
\end{figure}

We start with Lagrangian \eqref{Leff}, including only leader interactions which are relevant in $d=d_{uc}-\e$ with $\e\ll 1$:
\be\label{leadlag}
\Lcal_L^{V(\phi)}=\partial \vf \partial \omega-\frac H2 \omega^2 +\frac 12 \sum_{i=2}^n(\partial \chi_i)^2+  (\omega V'(\vf) + \chi_i^2V''(\vf))\,.
\ee 
We will work in the minimal subtraction (MS) scheme, so we dropped the mass term as usual. We may first set $V=0$ and write down the free theory propagators of different fields in momentum space. For what we discuss below we only need the following propagators explicitly:
\be
G_{\vf\vf}(p)=\frac{H}{p^4}\,, 
%\qquad G_{\omega\vf}(p)=\frac{1}{p^2}\,, 
\qquad G_{\chi_i\chi_j}(p)=\frac{K_{ij}}{p^2} \qquad \text{where}\ \  K_{ij}=\d_{ij}-\frac{\Pi_{ij}}{n-1}\,.
\ee
Here $\Pi_{ij}=1$ for all $i,j=2,\cdots , n$. %The matrix $K$ satisfies $K^2=K$\,, $\text{tr}\, K=n-2$\,, $\sum_i K_{ij}=0$. 
The factor $K_{ij}$ imposes the condition $\sum_{i=2}^n \chi_i=0$ from \eqref{Ctrans}.
In a Feynman diagram the propagators will be denoted as shown in Fig. \ref{Fig:propagators}. 

The propagator $G_{\vf\omega}$ is also present in the theory. It has the the same $1/p^2$ momentum dependence as $G_{\chi_i\chi_j}$ but we will not need it in the examples below.

We may now turn on interaction $V(\phi)=\frac{\l}{4!}\phi^4$ or $\frac{g}{6}\phi^3$ for which $d_{uc}=6$ or $8$ respectively.  Then we introduce bare and renormalized quantities (fields and coupling), and relate them by renormalization constants. These contants are obtained by requiring that correlators of renormalized quantities are finite as $\e\to 0$. 

We are interested in the $n\to0$ limit of the theory. Note that the $n\to0$ limit theory contains infinitely many fields $\chi_2,\chi_3,\chi_4,\ldots$ which can appear on the external legs. To take the $n\to 0$ limit of any Feynman diagram we have to simplify the product of matrices $K_{ij}$ from the $\chi$ propagators, using $\Pi_{ij} \Pi_{jk}= (n-1) \Pi_{ik}\to -\Pi_{ik}$. 

Consider first the $\phi^4$ case in $d=6-\e$. The one-loop beta function $\b_\l$ is computed from the coupling renormalization constant that can be obtained from a 4-point correlator e.g. $\langle \chi_i(p_1)\chi_j(p_2)\vf(p_3)\vf(p_4)\rangle$. Setting it to zero we get a fixed point at $\l_{\star}=\frac{64\pi^3\e}{3}$. We do not show the computations as the steps are very similar to the $d=4-\e$ usual $\phi^4$ (Wilson-Fisher) theory  (see e.g. \cite{Kleinert:2001ax}).

The field renormalizations are obtained from 2-point functions,  e.g. from the 2-loop correction to $\langle\chi_i(p)\chi_j(-p)\rangle$ one obtains the leading correction $\g_\chi$ to the dimension of $\chi_i$. We point out that due to the equivalence of \eqref{leadlag} as $n\to0$ with the SUSY theory the anomalous dimensions of all fundamental fields are equal, i.e. $\g_\vf=\g_\omega=\g_\chi=\frac{\e^2}{108}$. Note that this is the same field anomalous dimension as the usual Wilson-Fisher value, as expected from dimensional reduction. These computations are also very similar to $d=4-\e$ so we do not show them.

\begin{figure}
	\centering
	\hspace{-3cm}
	\begin{subfigure}[b]{0.6\textwidth}
		\centering
		\includegraphics[width=140pt]{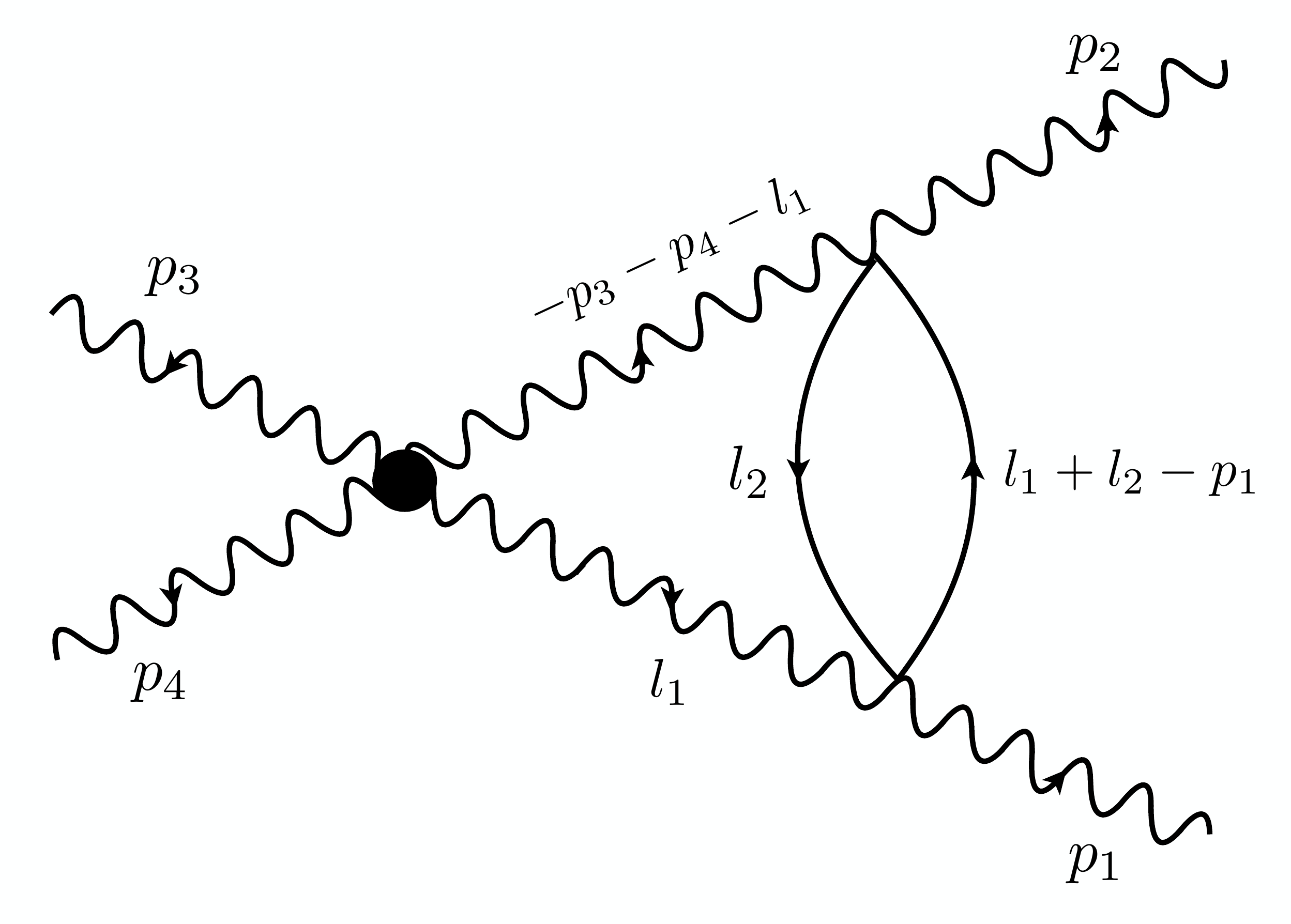}
		\caption{$V(\phi)=\frac{\l}{4!}\phi^4$,}
		\label{fig:2a}
	\end{subfigure}
	\hspace{0.1cm}
	\begin{subfigure}[b]{0.3\textwidth}
		\centering
		\includegraphics[width=140pt]{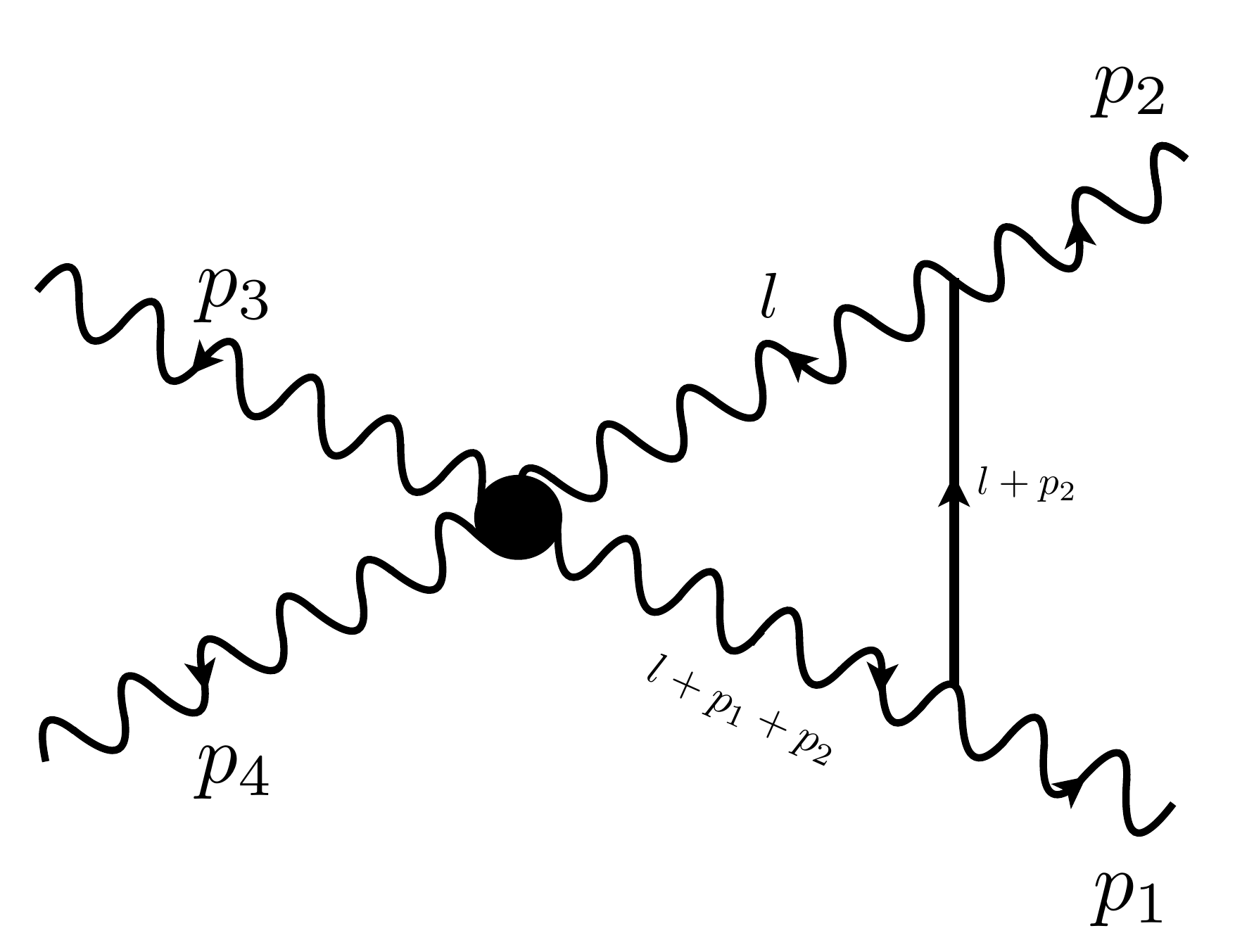}
		\caption{$V(\phi)=\frac{g}{6}\phi^3$.}
		\label{fig:2b}
	\end{subfigure}
	
	\caption{Computing the anomalous dimension of  $(\mathcal{N}_4)_L$ in the two theories.}
	\label{fig:2}
\end{figure}

Once we unerstand the RG flow of the basic Lagrangian \eqref{leadlag} we start perturbing it by other leader interactions. The couplings of those interactions are kept infinitesimal, as we are just interested to know their scaling dimension. In other words, we are computing anomalous dimensions of various local operators of the theory,
We do it in the standard way by defining the renormalization constant $Z_{\mathcal{O}}$ via $(\Ocal)^B=Z_{\Ocal} \Ocal$ where $(\Ocal)^B$ is a bare operator and $\Ocal$ a renormalized operator. Then we have the anomalous dimension $\g_{\Ocal}=\big[\frac{\partial}{\partial(\log \mu)} Z_{\Ocal}\big]_{\l=\l_{\star}}$.

Let us demonstrate the computation with the example of $(\mathcal{N}_4)_L=(\chi_i^2)^2$ which is the SN leader with the lowest classical dimension. We use the correlation function $\langle(\mathcal{N}_4)_L(p=0)\chi_i(p_1)\chi_j(p_2)\chi_k(p_3)\chi_l(p_4)\rangle$ and remove its $\eps\to0$ singularities to compute $Z_{(\mathcal{N}_4)_L}$. We choose this operator since its leading correction comes from a nontrivial 2-loop diagram shown in Fig. \ref{fig:2a}.

The  evaluation of this 2-loop integral is not uncommon in the $\phi^4$ literature \cite{Kleinert:2001ax}. The result is:
\be
\frac{H^2 \lambda^2}{(2 \pi)^{2 d}} \int \frac{d^d l_1
	d^d l_2}{l_1^2  (l_2^2)^2  (l_1 + p_3 + p_4)^2  ((l_1 + l_2 - p_1)^2)^2} =
\frac{H^2 \lambda^2}{2 (4 \pi)^6  \e} + O \left( \e^0 \right) .
\label{loopchi22}
\ee
Requiring that the $\e^{-1}$ divergence cancels (and taking into account the fundamental field renormalizations) we get:
\begin{equation}
Z_{(\mathcal{N}_4)_L}^{- 1} =1 - \frac{4}{3} \frac{H^2 \lambda^2}{(4 \pi)^6  \e} \implies
\gamma_{(\mathcal{N}_4)_L}  = -
\frac{8}{27} \e^2 . \label{gammachi4}
\end{equation}
All other operators presented in the main text that have a 2-loop leading anomalous dimension involve the same loop integral. For the ones with a 1-loop leading correction the computation is similar to that of the beta function.  The computations of beta function, field renormalization and anomalous dimensions of all operators considered in our work are shown in detail in App. H of \cite{paper2}.

For the $V(\phi)=\frac{g}{6}\phi^3$ potential in $d=8-\e$ we define the bare quantities and renormalization constants in a similar way. The beta function and field renormalization constants are obtained from e.g. the correlators $\langle\chi_i(p_1)\chi_j(p_2)\vf(p_3)\rangle$ and $\langle\chi_i(p_1)\chi_j(p_2)\rangle$ respectively. 
We get a fixed point at $g_\star^2=-\frac{2}{3}\frac{(4\pi)^4\e}{H}$\,. The field anomalous dimensions are $\g_\vf=\g_{\omega}=\g_{\chi} =-\frac{\e}{18}$. As expected these are same as the usual $\phi^3$ theory in $d=6-\e$ and the computations are also exactly similar (see e.g. \cite{srednicki_2007} for the usual $\phi^3$ theory literature). 

We may once again focus on the operator $(\mathcal{N}_4)_L$ and compute its anomalous dimension. In this case the leading correction comes from the diagram as shown in Fig. \ref{fig:2b}. The loop integral is similar to that of the beta function and computed using standard techniques of the usual $\phi^3$ theory. It gives
\be
\gamma_{(\mathcal{N}_4)_L}  = \frac{10}{9}\e\,.
\ee 
All other operators presented in the main text involve the same 1-loop integral. Details of all these computations will be given in a dedicated paper \cite{paper3}.

\end{document}